\documentclass[12pt,preprint]{aastex}

\slugcomment{ApJ in press}

\shorttitle{Stability of the Stalled Shocks
in Core-Collapse Supernovae}
\shortauthors{YAMASAKI \& YAMADA}

\begin{document}

\title{Stability of the Accretion Flows with Stalled Shocks \\
in Core-Collapse Supernovae}

\author{Tatsuya Yamasaki\altaffilmark{1} and Shoichi Yamada\altaffilmark{2,3}}

\altaffiltext{1}{Yukawa Institute for Theoretical Physics,
Kyoto University, Oiwake-cho, Sakyo, Kyoto 606-8502, Japan;
ytatsuya@yukawa.kyoto-u.ac.jp}

\altaffiltext{2}{Science and Engineering, Waseda University, 3-4-1 Okubo,
Shinjuku, Tokyo 169-8555, Japan;
shoichi@heap.phys.waseda.ac.jp}

\altaffiltext{3}{Advanced Research Institute for Science and Engineering,
Waseda University, 3-4-1 Okubo, Shinjuku, Tokyo 169-8555, Japan}

\begin{abstract}

Bearing in mind the application to the theory of core-collapse supernovae,
we performed a global linear analysis on the stability of spherically symmetric
accretion flows through a standing shock wave onto a proto neutron star.
As unperturbed flows, we adopted the spherically symmetric steady solutions
obtained
with realistic equation of state and formulae for neutrino reaction rates
taken into account.
These solutions are characterized by the mass accretion rate
and neutrino luminosity.
Then we solved the equations for linear perturbations numerically,
and obtained the eigen frequencies and eigen functions.
We found (1) the flows are stable for all modes
if the neutrino luminosity is lower than a certain value,
e.g. $\sim 1\times 10^{52}$ ergs/s
for $\dot{M}=1.0M_{\odot}/{\rm s}$.
(2) For larger luminosities, the non-radial instabilities
are induced, probably via the advection-acoustic cycles.
Interestingly, the modes with $\ell=2$ and $3$ become
unstable at first for relatively low neutrino luminosities,
e.g. $\sim 2-3\times 10^{52}$ ergs/s
for the same accretion rate, whereas the $\ell=1$ mode is
the most unstable for higher luminosities,
$\sim 3-7\times 10^{52}$ ergs/s.
These are all oscillatory modes.
(3) For still larger luminosities, $\gtrsim 7\times 10^{52}$ ergs/s
for $\dot{M}=1.0M_{\odot}/{\rm s}$,
non-oscillatory modes, both radial and non-radial, become unstable.
These non-radial modes were identified as convection.
The growth rates of the convective modes
have a peak at $\ell=5-11$, depending on the luminosity.
We confirmed the results obtained by numerical simulations that
the instabilities induced by the advection-acoustic cycles are
more important than the convection for lower neutrino luminosities.
Furthermore, we investigated the effects of
the inner boundary on the stabilities
and found that they are not negligible though the existence of
the instabilities is not changed qualitatively for a variety of conditions.

\end{abstract}


\section{Introduction}

The explosion mechanism of core-collapse supernova is still
remaining to be revealed. 
Several observations suggest that the supernova explosion is
asymmetric in general. 
For example, the HST images of supernova 1987A,
the most thoroughly investigated supernova so far,
shows clear global asymmetry of the ejecta.
The radiations scattered by the supernova ejecta 
commonly have one percent level of linear polarization,
the fact naturally explained by the asymmetry of
the scattering surface~\citep{leo00, wan02}.
Furthermore, it is now well known that young pulsars have
a large peculiar velocity \citep{lyn94}, which is supposed to be gained
by a nascent neutron star at its birth owing to the asymmetry of dynamics.
On the theoretical side, the possible roles of the asymmetry
in the explosion mechanism are being explored from various points of view
at the moment~\citep{kot06}. 

As a result of many multi-dimensional studies, it is now believed that
the dynamics of supernova core is intrinsically asymmetric 
even though extrinsic factors like the rotation or
magnetic field of progenitor is absent.
The asymmetry is thought to be produced by some hydrodynamical instabilities, 
the most well known of which is convection.
A couple of recent numerical simulations demonstrate, on the other hand, 
that another instability plays an important role in the accretion flows
through the standing shock 
onto the proto neutron star \citep{blo03,sch04,blo06,ohn06a}.
The remarkable feature of this instability is
the dominance of the $\ell=1$ mode, 
where $\ell$ stands for the azimuthal index of
the spherical harmonic functions, $Y_{\ell}^{m}$.
The so-called standing accretion shock instability or SASI is thus expected
to be a natural explanation for the large kick velocities of young pulsars. 

The mechanism of SASI is still controversial.
One of the promising mechanisms is the amplified 
cycle of the inward advections of vortex and entropy fluctuations
and outward propagations of acoustic perturbations,
which was originally discussed
in the context of black hole accretions \citep{fog00,fog01,fog02}.
The mechanism was also known in the engineering field
(see \citet{how75} and references therein).
Most recently, however, \citet{blo06} claimed, based on
their 2D numerical simulations with a simplified cooling
taken into account, that the instability is induced
by the purely acoustic cycles. 
On the other hand, \citet{ohn06a} obtained the results that
appear to support the advection-acoustic
cycle in their 2D numerical models with more realistic heating and cooling
processes taken into consideration. Obviously,
more detailed linear analyses for appropriate background
models and boundary conditions are needed to clarify
the nature of the instabilities and that is the main 
purpose of this paper, in which we accomplished global linear analyses
for the spherically symmetric steady accretion flows
that are supposed to represent the post-bounce supernova core reasonably well. 
It should be emphasized that SASI is a non-local instability.

As mentioned above, the convection has been numerically explored
by many researchers for the past decades. 
If the neutrino luminosity is larger than a certain value,
there appears a so-called heating region in the post-shock flows,
where the net neutrino heating occurs~\citep{yam06}.
In this region, the entropy gradient is negative and the flow is
convectively unstable if the conventional criterion for convection is applied. 
Recently, \citet{fog06a} demonstrated by linear analysis that
the advection tends to suppress the convection. 
In fact, the condition for the instability is tightened
and the growth rate is lowered in general.
In this paper, SASI was not discussed.
We think, however, the SASI and convection should be treated
simultaneously and on an equal basis.
It is also noted that it is important to employ a realistic equation of state
and neutrino heating/cooling rates, 
since not only the evolution of perturbations but also
the unperturbed background structure is dictated by them. 
In fact, the stability strongly depends on the latter. 
For example, the radial distributions of sound velocity
and inflow velocity are crucially important for the advection-acoustic cycle.
The growth rate and stability criterion for convection,
on the other hand, depend not only on the entropy gradient
but also on the advection time of the flow as discussed
by \citet{fog06a}.
Moreover, the stability of accretion flows through a standing shock
against radial perturbations is mainly determined
by the acceleration or deceleration of the flow just behind the shock wave
\citep{nak92,nak93,nak94}.
It is, hence, important to employ appropriate unperturbed background flows and
boundary conditions. Some of the discrepancy mentioned
above may be attributed to the difference of them. 

In this paper, we accomplished global linear analyses of
spherically symmetric steady accretion flows through a standing 
accretion shock onto a proto neutron star,
using a realistic equation of state and neutrino-heating/cooling rates, and 
investigated the stability of the flows against both radial
and non-radial perturbations. In so doing, we treated
the convective instabilities and SASI on an equal basis.
In the next section, we summarize the formulations, describing briefly 
the background steady solutions and then giving basic equations
for the global linear analysis.
The main results of the numerical calculations are
presented in section~\ref{eigen}.
We discuss in section~\ref{mec} the mechanisms of instabilities found
in the previous section.
In section~\ref{dep}, we consider the dependence of
the results on the assumptions we make for the inner boundary conditions
and the perturbation of neutrino temperature.
The final section is devoted to discussion.

\section{Formulations\label{form}}

\subsection{Unperturbed Steady Accretion Flows} \label{steady}

The unperturbed steady background flows were obtained exactly
the same way as in the previous paper \citep{yam06}.
The results are supposed to approximate the post-bounce accretion flows
through a stalled shock wave onto a proto neutron star in the supernova core.
The approximation is justified by the fact that the typical time scale
for the variations of mass accretion rate and neutrino luminosity is longer
than the dynamical time scale of the accretion flow.
We solved the hydrodynamical equations only for the region
between the shock front and the neutrino sphere,
approximating the upstream flow of the shock by the free fall.
We refer readers to \citet{yam06} for the basic equations (Eqs.~(1)-(5))
and other details, and summarize only some important elements here. 
The notations were also inherited from the paper.

The formulations are Newtonian.
We took account of the self-gravity of the accretion flow.
The neutrino transfer was not solved,
since the region of interest is optically thin for neutrinos. 
The neutrino luminosities and spectra were assumed to be constant in time. 
The latter were assumed to be thermal with a temperature of $4.5$MeV.
The realistic equation of state by \citet{she98} and
the standard reaction rates for the absorptions and emissions of 
neutrino on free nucleons by \citet{bru85} were employed.
The formulae for the latter we adopted in this paper are given in Appendix 
for later convenience.
As for the boundary conditions, we imposed the Rankine-Hugoniot relations
at the shock front.
At the inner boundary, on the other hand, the condition that
the optical depth for neutrino be $2/3$ was imposed.
See Eqs.~(6)-(10) and (17) in \citet{yam06}. 

We solved these equations for a wide range of the control parameters, that is, 
the mass accretion rate and neutrino luminosity.
The results were shown in Figs.~2-4 in \citet{yam06}. 
It is just repeated that there is a critical luminosity
for each given mass accretion rate, above which there exists no steady solution.

\subsection{Basic Equations for Perturbations} \label{equations}

The formulation is essentially based on the spherical coordinates
($r$, $\theta$, $\phi$). 
In order to treat a variable outer boundary position, however, we adopt a new 
variable $x$ instead of $r$ as follows,
\begin{equation}
x=\frac{r-r_{\nu}}{r_{\rm s}-r_{\nu}},
\label{eq1}
\end{equation}
where $r_{\rm s}$ and $r_{\nu}$ denote, respectively,
the radii of the shock front and of the inner boundary,
which corresponds to the neutrino sphere.
The time-dependent Euler equations are expressed as 
\begin{eqnarray}
\lefteqn{\frac{\partial \rho}{\partial t}
-\frac{x}{r_{\rm s}-r_{\nu}}
\frac{\partial r_{\rm s}}{\partial t}
\frac{\partial \rho}{\partial x}
+\frac{\rho}{r_{\rm s}-r_{\nu}}
\frac{\partial u_r}{\partial x}
+\frac{u_r}{r_{\rm s}-r_{\nu}}
\frac{\partial \rho}{\partial x}} \nonumber\\
& &+\frac{2\rho u_r}{r}
+\frac{\rho}{r}\frac{\partial u_{\theta}}{\partial \theta}
+\frac{\cot \theta \rho u_{\theta}}{r}
+\frac{\rho}{r\sin \theta}\frac{\partial u_{\phi}}{\partial \phi}
=0,
\label{eq2}
\end{eqnarray}
\begin{equation}
\frac{\partial u_r}{\partial t}
-\frac{x}{r_{\rm s}-r_{\nu}}
\frac{\partial r_{\rm s}}{\partial t}
\frac{\partial u_r}{\partial x}
+\frac{u_r}{r_{\rm s}-r_{\nu}}
\frac{\partial u_r}{\partial x}
+\frac{1}{r_{\rm s}-r_{\nu}}\frac{1}{\rho}\frac{\partial p}{\partial x}
+\frac{GM}{r^2}=0,
\label{eq3}
\end{equation}
\begin{equation}
\frac{\partial u_{\theta}}{\partial t}
+\frac{u_r}{r_{\rm s}-r_{\nu}}
\frac{\partial u_{\theta}}{\partial x}
+\frac{u_r u_{\theta}}{r}
+\frac{1}{r\rho}\frac{\partial p}{\partial \theta}
-\frac{1}{r\rho}\frac{x}{r_{\rm s}-r_{\nu}}
\frac{\partial r_{\rm s}}{\partial \theta}\frac{\partial p}{\partial x}=0,
\label{eq4}
\end{equation}
\begin{equation}
\frac{\partial u_{\phi}}{\partial t}
+\frac{u_r}{r_{\rm s}-r_{\nu}}
\frac{\partial u_{\phi}}{\partial x}
+\frac{u_r u_{\phi}}{r}
+\frac{1}{r\rho}\frac{\partial p}{\partial \phi}
-\frac{1}{r\rho}\frac{x}{r_{\rm s}-r_{\nu}}
\frac{\partial r_{\rm s}}{\partial \phi}\frac{\partial p}{\partial x}=0,
\label{eq5}
\end{equation}
\begin{equation}
\frac{\partial S}{\partial t}
-\frac{x}{r_{\rm s}-r_{\nu}}
\frac{\partial r_{\rm s}}{\partial t}
\frac{\partial S}{\partial x}
+\frac{u_r}{r_{\rm s}-r_{\nu}} \frac{\partial S}{\partial x}
=\frac{\dot{q}}{\rho T},
\label{eq6}
\end{equation}
\begin{equation}
\frac{\partial Y_e}{\partial t}
-\frac{x}{r_{\rm s}-r_{\nu}}
\frac{\partial r_{\rm s}}{\partial t}
\frac{\partial Y_e}{\partial x}
+\frac{u_r}{r_{\rm s}-r_{\nu}} \frac{\partial Y_e}{\partial x}
=\frac{\lambda}{n},
\label{eq7}
\end{equation}
where $u_r$, $u_{\theta}$, $u_{\phi}$ stand for the velocity components and 
$\rho$, $p$, $S$, $Y_e$ and $n$ represent the density,
pressure, entropy per unit mass, electron fraction and baryon number density,
respectively; $G$ is the gravitational constant and $\dot{M}$ is
the mass accretion rate, which is also constant in time; $M$ is
the enclosed mass; $\dot{q}$ and $\lambda$ are the net heating
and reaction rates for the absorptions and emissions of neutrinos
on free nucleons. 
The terms originating from the non-orthogonality of the new coordinate system are dropped here 
because they are of higher order in terms of the deviation
from the steady solution.

The formulation is again Newtonian and the rotation and magnetic field
were neglected.
Since the perturbed flows are also optically thin for neutrinos
in the region of interest, the neutrino transport was ignored.
In addition, we neglected the perturbations of the self-gravity
of the accretion flows. This is justified from the fact that
the inclusion of the self-gravity did not change much the unperturbed
steady solutions except for the case with the neutrino luminosity
very close to the critical value (see figure 3b in \citet{yam06}).
As will become clear later, the latter case is of no importance in reality.

The Rankine-Hugoniot relations at the outer boundary are expressed as
\begin{equation}
\rho({\bf u} -{\bf u}_{\rm s})\cdot {\bf n}
=\rho_{\rm u}({\bf u}_{\rm u} -{\bf u}_{\rm s})\cdot {\bf n},
\label{eq8}
\end{equation}
\begin{equation}
({\bf u} -{\bf u}_{\rm s})\cdot {\bf t}_{\alpha}
=({\bf u}_{\rm u} -{\bf u}_{\rm s})\cdot {\bf t}_{\alpha},
\label{eq9}
\end{equation}
\begin{equation}
({\bf u} -{\bf u}_{\rm s})\cdot {\bf t}_{\beta}
=({\bf u}_{\rm u} -{\bf u}_{\rm s})\cdot {\bf t}_{\beta},
\label{eq10}
\end{equation}
\begin{equation}
\rho\left\{({\bf u} -{\bf u}_{\rm s})\cdot \frac{\bf n}{|{\bf n}|}\right\}^2 +p
=\rho_{\rm u}\left\{({\bf u}_{\rm u} -{\bf u}_{\rm s})\cdot
\frac{\bf n}{|{\bf n}|}\right\}^2
+p_{\rm u},
\label{eq11}
\end{equation}
\begin{equation}
\frac{1}{2}\left\{({\bf u} -{\bf u}_{\rm s})\cdot
\frac{\bf n}{|{\bf n}|}\right\}^2
+\epsilon+\frac{p}{\rho}
=\frac{1}{2}\left\{({\bf u}_{\rm u} -{\bf u}_{\rm s})\cdot
\frac{\bf n}{|{\bf n}|}\right\}^2
+\epsilon_{\rm u}+\frac{p_{\rm u}}{\rho_{\rm u}},
\label{eq12}
\end{equation}
where ${\bf u}_{\rm s}$ and $\epsilon$ are the shock velocity and
specific internal energy, respectively;
The suffix ${\rm u}$ stands for the quantities in the upstream,
which was assumed to be free fall and cold; The vectors normal and tangent
to the shock surface are denoted as
${\bf n}$ and ${\bf t}_{\alpha}$, ${\bf t}_{\beta}$, respectively, 
and are given as 
\begin{equation}
{\bf n}=\left(1,-\frac{1}{r_{\rm s}}\frac{\partial r_{\rm s}}{\partial \theta},
-\frac{1}{r_{\rm s}\sin\theta}\frac{\partial r_{\rm s}}{\partial \phi}\right),
\label{eq13}
\end{equation}
\begin{equation}
{\bf t}_{\alpha}=\left(\frac{\partial r_{\rm s}}{\partial \theta},
r_{\rm s},0\right),
\label{eq14}
\end{equation}
\begin{equation}
{\bf t}_{\beta}=\left(\frac{1}{\sin\theta}
\frac{\partial r_{\rm s}}{\partial \phi},0,r_{\rm s}\right).
\label{eq15}
\end{equation}

For the linear analysis, we employed the formulation based on
the Eulerian perturbation at the fixed $x$-coordinate.
The deviations from the unperturbed states are expanded with
the spherical harmonics, $Y_{\ell}^m(\theta,\phi)$,
and are assumed to have a time dependence of $\exp(\omega t)$.
Attaching the subscript $0$ to the unperturbed variables and
the subscript $1$ to the deviations from them of the perturbed counterparts,
we expressed, for example, the radial velocity as 
\begin{equation}
u_r=u_{r0}(x)+\sum_{\ell, m}u_{r1}(x)Y_{\ell}^m(\theta,\phi) \, e^{\omega t}.
\label{eq16}
\end{equation}
The exceptions are $u_{\theta}$ and $u_{\phi}$,
which were expanded with the vector spherical harmonics as
\begin{equation}
\left(\begin{array}{c}
u_{\theta} \\ u_{\phi}
\end{array} \right) =
\sum_{\ell,m}u_{h1}(x)
\left(\begin{array}{c}
\displaystyle{\frac{\partial Y_{\ell}^m(\theta,\phi)}{\partial \theta}}\\
\displaystyle{\frac{\partial Y_{\ell}^m(\theta,\phi)}{\sin\theta \, \partial \phi}}
\end{array} \right) \, e^{\omega t}.
\label{eq17}
\end{equation}

Then the basic equations describing the perturbations to
the lowest order are given for each $\ell$ as follows:
\begin{eqnarray}
\lefteqn{\frac{d(\rho_1/\rho_0)}{dx}
+\frac{d(u_{r1}/u_{r0})}{dx}
+\frac{r_{{\rm s}0}-r_{\nu}}{u_{r0}}\omega\frac{\rho_1}{\rho_0}
-\ell (\ell + 1)\frac{r_{{\rm s}0}-r_{\nu}}{r}\frac{u_{h1}}{u_{r0}}} \nonumber\\
& &-\left[\left(\frac{d\ln \rho_0}{dx}
+\frac{d\ln |u_{r0}|}{dx}\right)
\frac{r_{{\rm s}0}}{r_{{\rm s}0}-r_{\nu}}
+\left\{\frac{\omega}{u_{r0}}\frac{d\ln \rho_0}{dx}
+\frac{2(r_{{\rm s}0}-r_{\nu})}{r^2}\right\}xr_{{\rm s}0}\right]
\frac{r_{{\rm s}1}}{r_{{\rm s}0}}=0,
\label{eq18}
\end{eqnarray}
\begin{eqnarray}
\lefteqn{\frac{d(u_{r1}/u_{r0})}{dx}
+\frac{p_0}{u_{r0}^2\rho_0}
\left(\frac{\partial \ln p}{\partial \ln \rho}\right)_0
\frac{d(\rho_1/\rho_0)}{dx}
+\frac{p_0}{u_{r0}^2\rho_0}
\left(\frac{\partial \ln p}{\partial \ln S}\right)_0
\frac{d(S_1/S_0)}{dx}
+\frac{p_0}{u_{r0}^2\rho_0}
\left(\frac{\partial \ln p}{\partial \ln Y_e}\right)_0
\frac{d(Y_{e1}/Y_{e0})}{dx}} \nonumber\\
& &+\left(\frac{r_{{\rm s}0}-r_{\nu}}{u_{r0}}\omega
+2\frac{d\ln |u_{r0}|}{dx}\right)\frac{u_{r1}}{u_{r0}}
+\left\{\frac{d\ln \rho_0}{dx}
\frac{\rho_0^2}{p_0}\left(\frac{\partial^2 p}{\partial \rho^2}\right)_0
-\left(\frac{\partial \ln p}{\partial \ln S}\right)_0
\frac{d\ln S_0}{dx}\right. \nonumber\\
& &+\left.\frac{d\ln S_0}{dx}
\frac{\rho_0 S_0}{p_0}
\left(\frac{\partial^2 p}{\partial \rho \partial S}\right)_0
-\left(\frac{\partial \ln p}{\partial \ln Y_e}\right)_0
\frac{d\ln Y_{e0}}{dx}
+\frac{d\ln Y_{e0}}{dx}\frac{\rho_0 Y_{e0}}{p_0}
\left(\frac{\partial^2 p}{\partial \rho \partial Y_e}\right)_0 \right\}
\frac{p_0}{u_{r0}^2\rho_0}\frac{\rho_1}{\rho_0} \nonumber\\
& &+\left\{\frac{d\ln \rho_0}{dx}
\frac{\rho_0 S_0}{p_0}
\left(\frac{\partial^2 p}{\partial \rho \partial S}\right)_0
+\left(\frac{\partial \ln p}{\partial \ln S}\right)_0
\frac{d\ln S_0}{dx}\right.
+\left.\frac{d\ln S_0}{dx}
\frac{S_0^2}{p_0}
\left(\frac{\partial^2 p}{\partial S^2}\right)_0 \right.\nonumber\\
& &+\left.\frac{d\ln Y_{e0}}{dx}\frac{S_0 Y_{e0}}{p_0}
\left(\frac{\partial^2 p}{\partial S \partial Y_e}\right)_0\right\}
\frac{p_0}{u_{r0}^2\rho_0}\frac{S_1}{S_0} \nonumber\\
& &+\left\{\frac{d\ln \rho_0}{dx}
\frac{\rho_0 Y_{e0}}{p_0}
\left(\frac{\partial^2 p}{\partial \rho \partial Y_e}\right)_0
+\frac{d\ln S_0}{dx}\frac{S_0 Y_{e0}}{p_0}
\left(\frac{\partial^2 p}{\partial S\partial Y_e}\right)_0
+\left(\frac{\partial \ln p}{\partial \ln Y_e}\right)_0
\frac{d\ln Y_{e0}}{dx} \right. \nonumber\\
& &\left.+\frac{d\ln Y_{e0}}{dx}\frac{Y_{e0}^2}{p_0}
\left(\frac{\partial^2 p}{\partial Y_e^2}\right)_0\right\}
\frac{p_0}{u_{r0}^2\rho_0}\frac{Y_{e1}}{Y_{e0}} \nonumber\\
& &-\left\{\frac{\omega}{u_{r0}}\frac{d\ln |u_{r0}|}{dx}
+\frac{2(r_{{\rm s}0}-r_{\nu})GM}{r^3 u_{r0}^2}\right\}
xr_{{\rm s}0}\frac{r_{{\rm s}1}}{r_{{\rm s}0}}\nonumber\\
& &-\left[\frac{d\ln |u_{r0}|}{dx}
+\left\{\left(\frac{\partial \ln p}{\partial \ln \rho}\right)_0
\frac{d\ln \rho_0}{dx}
+\left(\frac{\partial \ln p}{\partial \ln S}\right)_0
\frac{d\ln S_0}{dx}
+\left(\frac{\partial \ln p}{\partial \ln Y_e}\right)_0
\frac{d\ln Y_{e0}}{dx}\right\}
\frac{p_0}{u_{r0}^2\rho_0}\right] \nonumber\\
& &\times\frac{r_{{\rm s}0}}{r_{{\rm s}0}-r_{\nu}}
\frac{r_{{\rm s}1}}{r_{{\rm s}0}}=0,
\label{eq19}
\end{eqnarray}
\begin{eqnarray}
\lefteqn{\frac{d(u_{h1}/u_{0})}{dx}
+\left(\frac{r_{{\rm s}0}-r_{\nu}}{u_{r0}}\omega
+\frac{r_{{\rm s}0}-r_{\nu}}{r}
+\frac{d\ln |u_{r0}|}{dx}\right)
\frac{u_{h1}}{u_{r0}}} \nonumber\\
& &+\frac{r_{{\rm s}0}-r_{\nu}}{r}\frac{p_0}{u_{r0}^2 \rho_0}
\left\{\left(\frac{\partial \ln p}{\partial \ln \rho}\right)_0
\frac{\rho_1}{\rho_0}
+\left(\frac{\partial \ln p}{\partial \ln S}\right)_0\frac{S_1}{S_0}
+\left(\frac{\partial \ln p}{\partial \ln Y_e}\right)_0
\frac{Y_{e1}}{Y_{e0}}\right\} \nonumber\\
& &-\frac{1}{r}\frac{p_0}{u_{r0}^2 \rho_0}\frac{d\ln p_0}{dx}
xr_{{\rm s}0}\frac{r_{{\rm s}1}}{r_{{\rm s}0}}=0
\quad\qquad(\mbox{for $\ell\ne0$}),
\label{eq20}
\end{eqnarray}
\begin{eqnarray}
\lefteqn{\frac{d(S_1/S_0)}{dx}
-\frac{r_{{\rm s}0}-r_{\nu}}{S_0 u_{r0} \rho_0 T_0}
\left\{\rho_0\left(\frac{\partial\dot{q}}{\partial \rho}\right)_0
-\left(\frac{\partial \ln T}{\partial \ln \rho}\right)_0\dot{q}_0
-\dot{q}_0\right\}\frac{\rho_1}{\rho_0}
+\frac{d\ln S_0}{dx}\frac{u_{r1}}{u_{r0}}} \nonumber\\
& &+\left[\frac{r_{{\rm s}0}-r_{\nu}}{u_{r0}}\omega
+\frac{d\ln S_0}{dx}
-\frac{r_{{\rm s}0}-r_{\nu}}{S_0 u_{r0} \rho_0 T_0}
\left\{S_0\left(\frac{\partial\dot{q}}{\partial S}\right)_0
-\left(\frac{\partial \ln T}{\partial \ln S}\right)_0\dot{q}_0\right\}\right]
\frac{S_1}{S_0} \nonumber\\
& &-\frac{r_{{\rm s}0}-r_{\nu}}{S_0 u_{r0} \rho_0 T_0}
\left\{Y_{e0}\left(\frac{\partial \dot{q}}{\partial Y_e}\right)_0
-\left(\frac{\partial \ln T}{\partial \ln Y_e}\right)_0\dot{q}_0\right\}
\frac{Y_{e1}}{Y_{e0}} \nonumber\\
& &-\left\{\frac{d\ln S_0}{dx}
\frac{r_{{\rm s}0}}{r_{{\rm s}0}-r_{\nu}}
+\left(\frac{\omega}{u_{r0}}\frac{d\ln S_0}{dx}
+\frac{r_{{\rm s}0}-r_{\nu}}{S_0 u_{r0} \rho_0 T_0}
\frac{\partial\dot{q}_0}{\partial r}\right)x r_{{\rm s}0}\right\}
\frac{r_{{\rm s}1}}{r_{{\rm s}0}} \nonumber\\
& &-\frac{r_{{\rm s}0}-r_{\nu}}{S_0 u_{r0} \rho_0 T_0}
T_{\nu 0}\left(\frac{\partial\dot{q}}{\partial T_{\nu}}\right)_0
\frac{T_{\nu 1}}{T_{\nu 0}}=0,
\label{eq21}
\end{eqnarray}
\begin{eqnarray}
\lefteqn{\frac{d(Y_{e1}/Y_{e0})}{dx}
+\frac{d\ln Y_{e0}}{dx}\frac{u_{r1}}{u_{r0}}
-\frac{r_{{\rm s}0}-r_{\nu}}{Y_{e0}n_0 u_{r0}}
\rho_0\left(\frac{\partial \lambda}{\partial \rho}\right)_0
\frac{\rho_1}{\rho_0}
-\frac{r_{{\rm s}0}-r_{\nu}}{Y_{e0}n_0 u_{r0}}
S_0\left(\frac{\partial \lambda}{\partial S}\right)_0
\frac{S_1}{S_0}} \nonumber\\
& &+\left\{\frac{r_{{\rm s}0}-r_{\nu}}{u_{r0}}\omega
+\frac{d\ln Y_{e0}}{dx}
-\frac{r_{{\rm s}0}-r_{\nu}}{Y_{e0}n_0 u_{r0}}
Y_{e0}\left(\frac{\partial \lambda}{\partial Y_e}\right)_0\right\}
\frac{Y_{e1}}{Y_{e0}} \nonumber\\
& &-\left\{\left(\frac{\omega}{u_{r0}}\frac{d\ln Y_{e0}}{dx}
+\frac{r_{{\rm s}0}-r_{\nu}}{Y_{e0}n_0 u_{r0}}
\frac{\partial \lambda_0}{\partial r}\right)xr_{{\rm s}0}
+\frac{d\ln Y_{e0}}{dx}
\frac{r_{{\rm s}0}}{r_{{\rm s}0}-r_{\nu}}\right\}
\frac{r_{{\rm s}1}}{r_{{\rm s}0}} \nonumber\\
& &-\frac{r_{{\rm s}0}-r_{\nu}}{Y_{e0}n_0 u_{r0}}T_{\nu 0}
\left(\frac{\partial\lambda}{\partial T_{\nu}}\right)_0
\frac{T_{\nu 1}}{T_{\nu 0}}
+\frac{r_{{\rm s}0}-r_{\nu}}{Y_{e0}n_0 u_{r0}}\lambda_0\frac{n_1}{n_0}=0,
\label{eq22}
\end{eqnarray}
where $T$ and $T_{\nu}$ denote the temperatures of matter and neutrino,
respectively.
The eigen values are degenerate in $m$
for the spherically symmetric background. 
The index $\ell$ explicitly appears only in Eq.~(\ref{eq18}). 

The Rankine-Hugoniot relations for the perturbed quantities
are written down similarly as
\begin{equation}
(2\rho_{{\rm f}0}u_{{\rm f}0}+\omega r_{{\rm s}0}\rho_{{\rm f}0}
-\omega r_{{\rm s}0}\rho_0 ) \frac{r_{{\rm s}1}}{r_{{\rm s}0}}
+\rho_0 u_{r0} \left(\frac{\rho_1}{\rho_0}+\frac{u_{r1}}{u_{r0}}\right)=0,
\label{eq23}
\end{equation}
\begin{equation}
\frac{u_{h1}}{u_{r0}}-\left(\frac{u_{{\rm f}0}}{u_{r0}}-1\right)
\frac{r_{{\rm s}1}}{r_{{\rm s}0}}=0
\quad\qquad(\mbox{for $\ell\ne0$}),
\label{eq24}
\end{equation}
\begin{eqnarray}
\lefteqn{\left\{\frac{5}{2}\rho_{{\rm f}0}u_{\rm f 0}^2
+2\omega r_{{\rm s}0}\rho_{{\rm f}0}u_{{\rm f}0}
-2\omega r_{{\rm s}0}\rho_0 u_{r0}
+\frac{3}{2}\rho_{{\rm f}0}
\left(\frac{\partial p}{\partial \rho}\right)_{{\rm f}0}\right\}
\frac{r_{{\rm s}1}}{r_{{\rm s}0}}} \nonumber\\
& &+2\rho_0 u_{0}^2 \frac{u_{r1}}{u_{r0}}
+\left\{\rho_0 u_{r0}^2 +\rho_0
\left(\frac{\partial p}{\partial \rho}\right)_0\right\}
\frac{\rho_1}{\rho_0}
+S_0\left(\frac{\partial p}{\partial S}\right)_0\frac{S_1}{S_0}
+Y_{e0}\left(\frac{\partial p}{\partial Y_e}\right)_0\frac{Y_{e1}}{Y_{e0}}=0,
\label{eq25}
\end{eqnarray}
\begin{eqnarray}
\lefteqn{\left\{\frac{1}{2}u_{{\rm f}0}^2
+\omega r_{{\rm s}0}u_{{\rm f}0}
-\omega r_{{\rm s}0}u_{r0}
+\frac{3}{2}\rho_{{\rm f}0}
\left(\frac{\partial \epsilon}{\partial \rho}\right)_{{\rm f}0}
+\frac{3}{2}\left(\frac{\partial p}{\partial \rho}\right)_{{\rm f}0}
-\frac{3}{2}\frac{p_{{\rm f}0}}{\rho_{{\rm f}0}}
\right\}\frac{r_{{\rm s}1}}{r_{{\rm s}0}}} \nonumber\\
& &+u_{r0}^2 \frac{u_{r1}}{u_{r0}}
+\left\{\rho_0\left(\frac{\partial \epsilon}{\partial \rho}\right)_0
+\left(\frac{\partial p}{\partial \rho}\right)_0
-\frac{p_0}{\rho_0}\right\}\frac{\rho_1}{\rho_0}
+\left\{S_0\left(\frac{\partial \epsilon}{\partial S}\right)_0
+\frac{S_0}{\rho_0}\left(\frac{\partial p}{\partial S}\right)_0\right\}
\frac{S_1}{S_0} \nonumber\\
& &+\left\{Y_{e0}\left(\frac{\partial \epsilon}{\partial Y_e}\right)_0
+\frac{Y_{e0}}{\rho_0}\left(\frac{\partial p}{\partial Y_e}\right)_0\right\}
\frac{Y_{e1}}{Y_{e0}}=0.
\label{eq26}
\end{eqnarray}
Note that the upstream flows are unaffected by the perturbations
inside the shock because they are supersonic and
no fluctuations can propagate beyond the shock front.
The suffix ${\rm f}0$ stands for the upstream values at the shock front
for the free fall assumed in the unperturbed background state.

In addition, we imposed 
\begin{equation}
u_{r1}=0,
\label{eq27}
\end{equation}
at the inner boundary ($x=0$), considering that the infalling matter
comes to rest on the neutron star surface that is supposed to be
very close to the inner boundary;
We did not try to obtain the "unsteady" structure between the inner boundary 
and the neutron star surface where a thin boundary layer is formed,
but chose instead to impose the boundary conditions
for perturbations at the neutrino sphere although $u_{r0}\ne0$ there,
although we must impose the boundary condition at the top
of the thin boundary layer between accretion flow
and the neutron star surface, in principle.
It is mentioned that as inferred from
the detailed numerical simulations of
the structure near the neutrino sphere \citep{lie05},
the distance between the neutrino sphere and the boundary layer
is indeed much smaller than the typical wavelengths of
the eigen modes we are concerned with.
At the same time, the advection time from shock to the boundary layer
is hardly altered by replacing the surface to the neutrino sphere;
That is, the effect of the inner boundary is almost unchanged by the replacing.
The effect of shifting the inner boundary will be discussed in \S~\ref{dep}.
Further, since the growth of SASI is expected to be
affected by the imposed inner boundary conditions, we tried some variations of
these conditions, the results of which will be discussed
in \S~\ref{dep}.

We solved Eqs.~(\ref{eq18})-(\ref{eq22}) as an eigen value problem for $\omega$,
imposing the boundary conditions, Eqs.~(\ref{eq23})-(\ref{eq27}).
The procedure is as follows. We first take a trial value of $\omega$, set 
$r_{{\rm s}1}/r_{{\rm s}0}$ to be unity and solve
the Rankine-Hugoniot relations, Eqs.~(\ref{eq23})-(\ref{eq26}),
and integrate Eqs.~(\ref{eq18})-(\ref{eq22})
from the outer boundary down to the inner boundary.
The solution obtained this way does not, in general, meet the inner boundary
condition, Eq.~(\ref{eq27}).
Then we improve the value of $\omega$ and repeat the procedure
until Eq.~(\ref{eq27}) is satisfied at the inner boundary by the solution.

\section{Eigen Frequencies of Various Modes}\label{eigen}

In this section, we summarily present various modes that we found in this paper. 
The detailed analysis of each mode will be given in the following sections. 
We show only the results for $\dot{M}=1.0M_{\odot}/{\rm s}$,
since their quantitative nature is weakly dependent on
the value of mass accretion rate. Unlike in the previous paper \citep{yam05}, 
in this and next sections, the neutrino temperature 
is assumed to be unperturbed, that is, $T_{\nu 1}=0$.
This is also equivalent to the negligence of the perturbation
of neutrino luminosity. The underlying idea is that the 
neutrino luminosity is dominated by the contribution
from the proto neutron star rather than from the accreted matter.
The influence of the perturbation of neutrino luminosity will
be considered separately in section~\ref{dep}.
We explored all the modes of relevance below 
the critical neutrino luminosity,
which is $124.801\times 10^{51}{\rm ergs/s}$ for
the mass accretion rate given above. 

First, we consider the radial modes with $\ell=0$.
It is noted that the stability of spherical accretion flows
against radial modes were investigated by several authors
\citep{nak92,nak93,nak94,hou92,fog00,fog01,fog02,gal05,yam05}
in different contexts.
\citet{nak92,nak93,nak94}, for example, showed that
the isothermal or adiabatic transonic accretion flows around black hole are
unstable when the postshock flow is accelerated.
The effect of cooling was investigated by \citet{hou92}
for the accretion flows to neutron stars or white dwarfs  
with a use of a simple analytic form of cooling.
\citet{yam05} analyzed the linear stability of spherically symmetric subsonic
accretion flows from a standing shock down to a proto neutron star, using
an analytic form of heating and cooling by neutrinos taken into account. 
This paper is an extention of \citet{yam05} to the models
with more realistic neutrino heating/cooling and equation of state that are
appropriate for the post-bounce phase of core-collapse supernovae 
(see also \citet{yam06}).

\begin{figure}
\plottwo{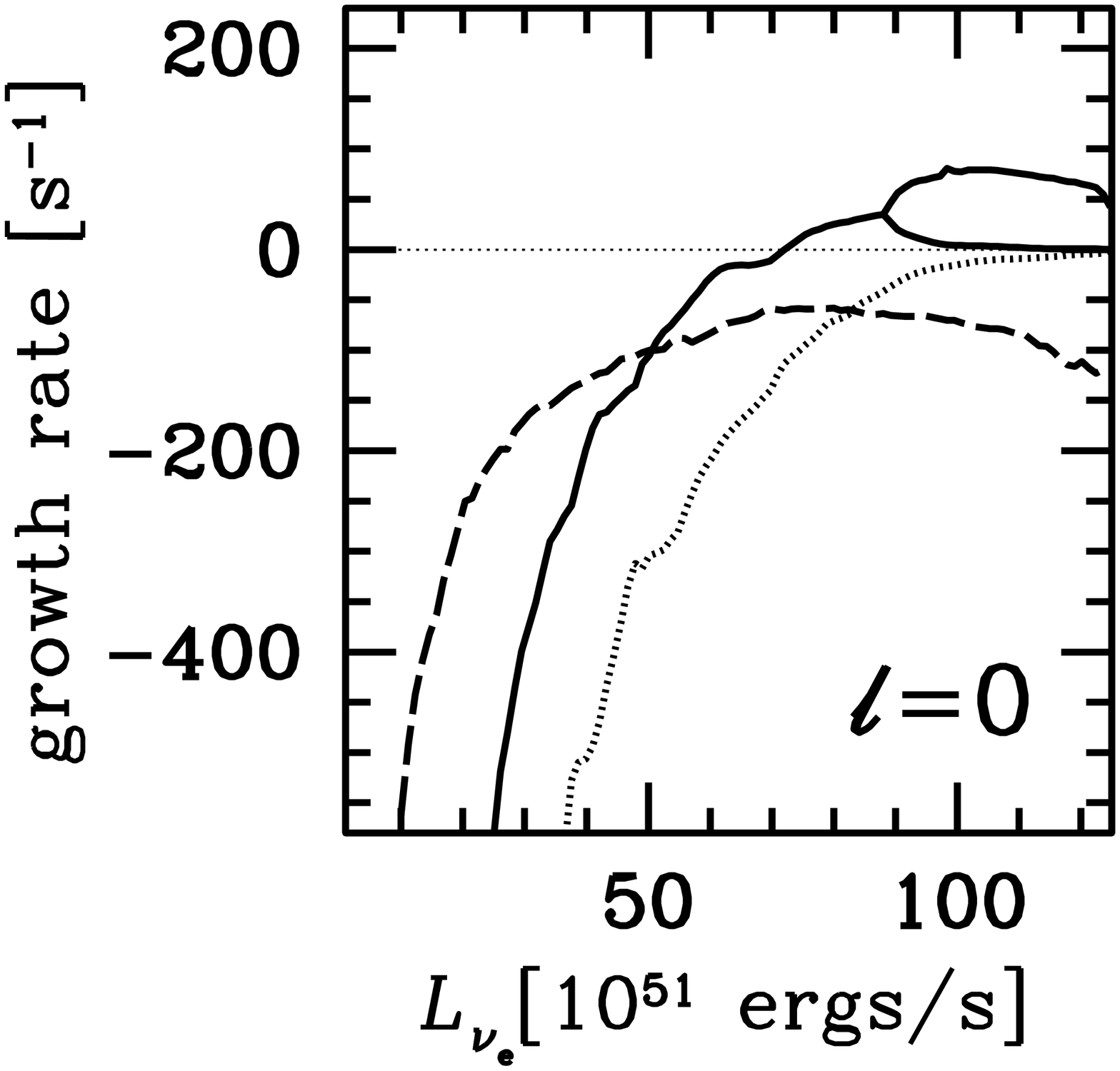}{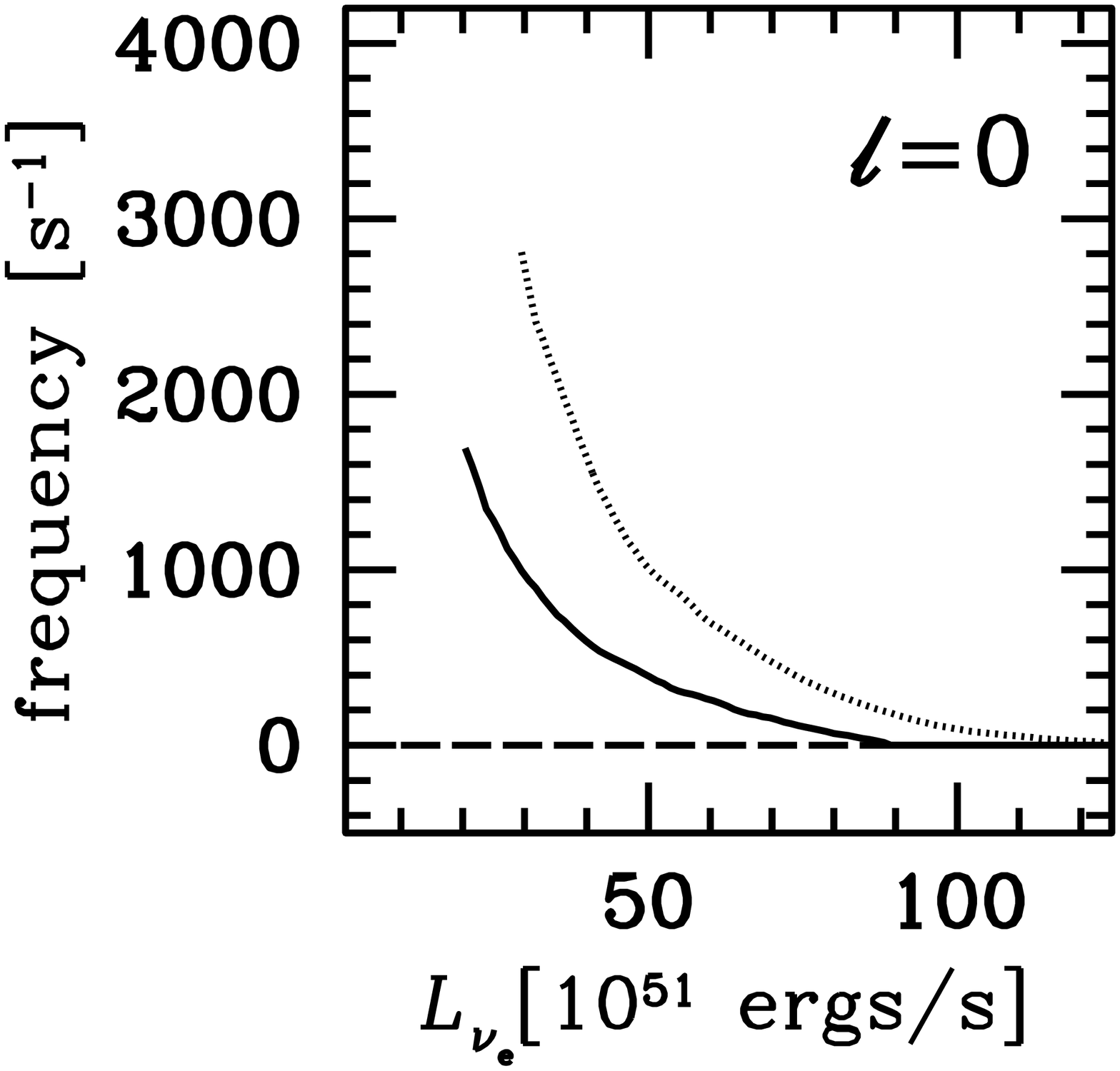}
\caption{{\it Left:} The growth rates, ${\rm Re}(\omega)$,
and {\it Right:} oscillation
frequencies, ${\rm Im}(\omega)$, for representative radial ($\ell=0$) modes
as a function of neutrino luminosity.
The long dashed curve denotes the fundamental thermal mode.
The solid and dotted curves correspond to 
the fundamental and first overtone oscillatory modes, respectively.
\label{fig1}}
\end{figure}

The results are shown in Fig.~\ref{fig1},
where the real and imaginary parts of $\omega$ (see Eq.~(\ref{eq16})) 
are displayed for representative radial modes.
It is found from the right panel that there are both oscillatory
(Im($\omega$)$\ne0$) and non-oscillatory (Im($\omega$)$=0$) modes. 
The mode shown with a long-dashed line is non-oscillatory and stable
(Re($\omega$)$<0$) for all the luminosities.
We distinguish the eigen modes with similar features
by the number of radial nodes.
We refer to the eigen mode with the smallest number of radial nodes,
normally none or one, as a fundamental mode of the family.
The modes with the second and third smallest number of radial nodes are called
the first and second overtones, respectively. In this notation,
the above mentioned mode is a fundamental mode. 
This mode is of thermal nature in the sense that
it disappears if the perturbations of the heating rates are neglected,
and has no non-radial counterpart.
This corresponds to the mode that was discussed in \citet{yam05}.
It is interesting to point out that the mode is not unstable even marginally at the critical luminosity.
This issue will be discussed further in detail in \S \ref{dep}.

It is apparent from Fig.~\ref{fig1} that the radial stability is
much more complicated in the present models than in those considered
in \citet{yam05}.
In fact, besides the thermal modes just discussed, we found oscillatory modes
shown with solid and dashed lines in the figure.
These are the fundamental and first overtone modes. This complexity
is originated from the non-monotonic background flows obtained
for the realistic neutrino heating rates and equation of state.
We refer readers to \citet{yam06} for more details on the unperturbed states.
While the first overtone (the dotted line) is stable and
oscillatory for all the luminosities up to the critical value, 
the fundamental mode becomes unstable when the luminosity is larger than
$\sim 7\times 10^{52}{\rm ergs/s}$.
For the luminosities greater than $\sim 9\times 10^{52}{\rm ergs/s}$,
the mode bifurcates into two branches and become non-oscillatory.
Although both the non-oscillatory modes are 
unstable, one of them is almost neutral for most of the luminosities.
The overstablizing radial mode that exists
for the luminosities $\lesssim 9\times 10^{52}{\rm ergs/s}$ 
was not found in \citet{yam05} but observed 
in the numerical simulations by \citet{ohn06a} and the reason was not 
clear in the latter paper. (Note, however, that such overstablizing modes were found 
in their own models by \citet{hou92}, \citet{gal05} and \citet{blo06}.)

We can now ascribe the apparent discrepancy to the difference of
the implemented microphysics. 
In fact, we re-examined the models with the simplified neutrino-heating rates
and equation of state employed in \citet{yam05} and found that
the same mode exists in this case also indeed, but it is stable up to
the critical luminosity.
In \citet{yam05}, the critical luminosity was reached
before the present mode becomes unstable.
In other words, which mode becomes unstable for smaller neutrino luminosities,
the present mode or the thermal mode, is dependent on the background model,
which is in turn determined by the adopted microphysics.
The oscillatory instability seems to be driven
by the advection-acoustic cycles \citep{fog00,fog01,fog02}. 
As shown in Fig.~\ref{fig4} together with other non-radial modes
that will be discussed later, the oscillation frequency agrees
with the period of the advection-acoustic cycle rather well. 
As will be described shortly, the non-radial modes become unstable
for smaller neutrino luminosities than these radial unstable modes.
Hence the latter modes will be of little practical importance. 
Other than the modes mentioned above, we did not find any unstable radial modes.

Next we pay attention to the stability of accretion flows
against non-radial perturbations.
\citet{fog00,fog01,fog02} conducted linear analyses
on the stability of black hole accretions, using a polytropic equation of state.
They claimed that the origin of the instability is
a repeated coupling of inward-advecting vortices and
entropy perturbations with outward-propagating acoustic waves.
It should be noted that the results cannot be directly applied
to the post-bounce flows in the supernova core of our current interest.
While in the the black hole accretion the flow is transonic
and the inner boundary is the event horizon,
in the latter case the flow is subsonic and
there is a surface of proto neutron star at the inner boundary.
\citet{gal05} attempted a linear analysis more appropriate for
the supernova core but with a polytopic equation of state
and a simple analytic cooling term.
Several numerical studies were also performed \citep{blo03,sch04,blo06,ohn06a},
and they all found the growths of unstable modes with $\ell=1$ and $2$.

The results of our calculations are shown in Figs.~\ref{fig2} and \ref{fig3},
where the growth rates and oscillation frequencies of the fundamental 
and first three overtone modes are given for $\ell=1-4$.
All the modes are oscillatory ($\rm{Im}(\omega) \ne 0$) 
for small neutrino luminosities $\lesssim 7\times 10^{52}{\rm ergs/s}$.
The accretion flows are stable against all the non-radial modes
if the neutrino luminosity is  
$\lesssim 1\times 10^{52}$ ergs/s for the present model with
$\dot{M}=1.0M_{\odot}/{\rm s}$.
For larger luminosities, the first overtones of $\ell=2$ and $3$ modes
become unstable at first for relatively low neutrino luminosities, 
$2 \sim 3\times 10^{52}$ ergs/s, whereas the fundamental modes with
$\ell=1$ and $2$ become unstable for a bit higher luminosities,
$3 \sim 7\times 10^{52}$ ergs/s, with the former more unstable.
For $\ell>3$, there is no unstable oscillatory mode.
For still larger luminosities, $\gtrsim 7\times 10^{52}$ ergs/s,
the fundamental modes of all $\ell$'s become non-oscillatory
($\rm{Im}(\omega) = 0$) and unstable.
It is interesting that they split into two different modes at the
same neutrino luminosity.
These features are quite similar to the radial fundamental modes, but are 
sensitive to the treatment of neutrino-luminositiy perturbations,
which will be discussed in \S \ref{dep}. 

\section{Mechanisms of Instabilities}\label{mec}

We now discuss the nature of each mode more in detail.
In the previous section, we found both oscillatory and non-oscillatory modes.
In the following, we treat them separately. 

\subsection{Oscillatory Modes --- Advection-Acoustic Cycle
or Purely Acoustic Cycle?} \label{oscillatory}

As shown in the previous section, all the modes except for the radial
fundamental thermal mode are oscillatory for low neutrino luminosities.
Although most of them are decaying oscillators, some of them,
e.g. the fundamental modes of $\ell=0, 1$ and the first overtones of
$\ell=2, 3$ modes, become growing oscillators
at a certain neutrino luminosity that is specific to each mode.
Recently, \citet{blo06} and \citet{ohn06a} did 2D numerical experiments
on the instability of the accretion flows onto a proto neutron star
and observed the non-radial oscillatory instabilities.
Moreover, they found the oscillating radial mode
that had not been expected in the paper by \citet{yam05}
as mentioned in \S \ref{eigen}.
Although the oscillation frequencies were consistent with each other,
these authors preferred a different interpretation on the mechanism
of the instabilities.
While \citet{ohn06a} supported the advection-acoustic-cycle mechanism
proposed by \citet{fog00,fog01}, \citet{blo06} claimed that
the purely acoustic cycles were responsible for the instabilities. 
In order to shed light on this issue,
we pay attention in this section to the possible origin of the oscillations. 

We first compare a couple of frequencies of relevance. 
We define the time scales of the advection-acoustic cycle, $\tau_{\rm aa}$, and
purely acoustic cycle, $\tau_{\rm pa}$, along the radial path as follows,
\begin{equation}
\tau_{\rm aa}\equiv\int^{r_{\rm s}}_{r_{\nu}}
\left(\frac{1}{c_s+u_r}-\frac{1}{u_r}\right)dr,
\label{eq30}
\end{equation}
\begin{equation}
\tau_{\rm pa}\equiv\int^{r_{\rm s}}_{r_{\nu}}
\left(\frac{1}{c_s+u_r}+\frac{1}{c_s-u_r}\right)dr,
\label{eq31}
\end{equation}
where $c_s$ is the sound speed. These frequencies are displayed together
with the oscillation frequencies obtained for various modes in Fig.~\ref{fig4}.
We can see that $2\pi/\tau_{\rm aa}$, the characteristic frequency of the
advection-acoustic cycle, agrees rather well with the oscillation frequencies
of the radial and non-radial modes 
whereas the characteristic frequency for the acoustic cycle
along the radial path, $2\pi/\tau_{\rm pa}$, is much larger. 
In fact, $\tau_{\rm aa}$ is almost coincident with the oscillation frequency
of the radial oscillatory mode.
\citet{blo06} insisted that the sound propagation is not radial
for $\ell \ne 0$ modes. For comparison, we also calculated 
the characteristic frequency for the sound wave to travel around
along the shock front although such a propagation of sound wave is highly
unlikely, since the radial gradient of the sound velocity is negative
and the sound wave will be deflected outwards like seismic waves.
This frequency is in better agreement with the oscillation frequencies than 
$\tau_{\rm pa}$, but appears to be too small compared with
the oscillation frequencies for low neutrino luminosities, 
particularly when the modes are unstable.
For these low luminosities, the shock front is close to the inner boundary 
and the radial path is much shorter than the circular path, and, as a result,
the difference of the interpretations is expected  to be most remarkable.
Note that the round-trip frequency becomes closer to
the oscillation frequencies at the lowest luminosities.
This is due to the non-monotonic radial dependence of sound velocity
at these luminosities, where the vicinity of the shock front has the largest
sound speed. These results, although not conclusive, seem to support
the advection-acoustic-cycle origin of the oscillation frequencies

Next, we discuss whether the advection of entropy-
and velocity perturbations and the propagation of pressure 
fluctuations actually occur in our models. 
In Figs.~\ref{fig5} and \ref{fig6} shown are the eigen functions of
the oscillatory modes of $\ell=1, 2$.
In drawing these figures, we changed the radial coordinate from $x$ to $r$
so that the Eulerian perturbations
of variables for fixed $r$ could be obtained.
For example, this Eulerian perturbation of $u_r$ denoted by the prime 
is related with another Eulerian perturbation for fixed $x$
by the following relation.
\begin{equation}
u'_r \equiv u_{r1}-xr_{{\rm s}1}\frac{du_{r0}}{dr}.
\label{eq28}
\end{equation}
Note that the luminosities are chosen so that
the displayed modes are unstable, ${\rm Re}(\omega)>0$.
The eigen functions are complex quantities in general and
both the real and imaginary part are presented in the figures.

In order to see if a particular fluctuation is propagated inward or outward,
we calculated the "phase" of the perturbed 
quantity defined, for example, for the horizontal velocity as follows: 
\begin{equation}
\phi\equiv \arccos\left(\frac{{\rm Re}(u_{h1})}{|u_{h1}|}\right).
\label{eq29}
\end{equation}
The values within $[0, 2\pi]$ are chosen at the shock front
and extended continuously inwards. 

In Fig.~\ref{fig7} given are the phases of the perturbations of pressure,
horizontal velocity and entropy. 
We also present the corresponding amplitudes in Fig.~\ref{fig8}. 
It is clear from Fig.~\ref{fig7} that the phases of the perturbations of
the horizontal velocity and entropy increase with radius,
which implies that they are propagating inward.
(Note that the perturbation is proportional to
$\exp (i {\rm Im}(\omega) t)$).
On the other hand, the phase of the pressure perturbation is 
almost constant with radius.
There is a slight positive gradient in the inner region,
indicating the inward propagation there, and a negative gradient
in the outer region, implying the outward propagation.
It should be also mentioned that the wave length of the pressure perturbations,
that is the sound velocity times the oscillation frequency,
is longer than the distance between the shock wave and the inner boundary.
This is also reflected in the lack of short-wave-length modulations of
the amplitude for the pressure perturbation shown
in Figs.~\ref{fig5} and \ref{fig6}.
In such a situation, it may not be meaningful to define a propagation path of
the pressure perturbation and this might be the source of difficulty
in interpreting the mechanism of the instability.

>From Fig.~\ref{fig8}, we can see which inward-advected quantity,
the perturbation of entropy or horizontal velocity, could give 
a larger contribution to the excitation of pressure perturbations.
Note that the amplitudes are normalized at the shock front.
As is clear, the entropy perturbation is significantly damped
from the initial value by the time it reaches the inner boundary.
This is particularly the case for the fundamental mode of
$\ell=1$ at $L_{\nu_{\rm e}}=6\cdot 10^{52}$ergs/s. 
The evolution of the entropy perturbation is determined
by the non-adiabatic processes, that is the heating and cooling by neutrinos.
A similar damping of the entropy perturbation is also seen
for the first overtone  of $\ell=2$ mode at
$L_{\nu_{\rm e}}=3\cdot 10^{52}$ergs/s although the damping region is
rather confined to the vicinity of the inner boundary.
The perturbation of the horizontal velocity, on the other hand,
is not reduced significantly as it is advected inward regardless of 
the unstable modes. Hence, if it is the advection-acoustic cycle that
drives the instability, the vortices rather than the 
entropy fluctuations will be the dominant source of the excitation of
pressure perturbations. It is incidentally
mentioned that the pressure perturbation is also damped rather
rapidly as it propagates outward as shown in Fig.~\ref{fig8}.

It should be emphasized that pressure perturbations are generated
at the inner boundary when vorteces arrive there if
the solid boundary condition is imposed~\citep{fog05}.
Since such pressure perturbations grow typically on the advection
time scale, it can seriously affect the advection-acoustic cycle excited 
by the pressure perturbations produced physically outside the inner boundary 
if the advection time is comparable to the growth time of the cycle.
The generations of pressure perturbations just on the inner boundary 
should be distinguished from those outside the boundary, the original 
excitation mechanism suggested by \citet{fog00,fog01,fog02}.
We cannot tell, however, to what extent the former contaminates 
the latter and how for the present cases from the above analyses alone.
We will discuss the issue in more detail in \S 5.

In our models meant for the core-collapse supernovae,
the unstable modes induced supposedly by the vortex-acoustic cycle 
have lower oscillation frequencies compared with the models meant for
the black hole accretion by \citet{fog01,fog02}.
This difference comes from the condition satisfied at the inner boundary,
around which the excitation of acoustic waves takes place most efficiently.
(See \citet{fog04} for the physical interpretation.)
Whereas the inflow velocity approaches the sound velocity at the inner boundary
in the case of the black hole accretion, the flow is highly subsonic near the
inner boundary in the present case, which lead to the observed lower frequencies
of the unstable oscillations induced by advected vortices in our models.

\subsection{Non-Oscillatory Modes --- Convective Instability}
\label{nonoscillatory}

As shown in the previous section, when the luminosity is
larger than about $7\times 10^{52}{\rm ergs/s}$,
there exit non-oscillatory unstable modes for all $\ell$'s including $\ell=0$. 
(see Figs.~\ref{fig1} and \ref{fig2}.)
Although the dependence of the growth rate
on the neutrino luminosity looks similar for these unstable modes, 
the eigen functions of the non-radial modes ($\ell\ne0$) are qualitatively 
different from that of the radial mode ($\ell=0$). 
This can be seen in  Fig.~\ref{fig9},
where the eigen functions of the modes with $\ell=0-4$ are
displayed and the gain radius is marked by the thin dotted lines.
While the eigen function of the entropy perturbation
for the radial mode has no node,
those for the non-radial modes have a node near the gain radius.
The eigen functions of the radial-velocity perturbation also show
a difference between them;
they take the maximum absolute value near the gain radius
for the non-radial modes,
whereas a rather monotonic decay is seen for the radial mode
as the radius decreases.
The fact that the heating region between the shock and
the gain radius has a negative entropy gradient and
that these unstable non-radial modes appear to be excited
in this region suggests that they are convective modes.
(The radial mode cannot be the convective mode in nature).

The growth rates of the convective modes together
with the radial mode are demonstrated in Fig.~\ref{fig10}. 
When the luminosity becomes greater than $\sim 7\times 10^{52}{\rm ergs/s}$,
all the modes are bifurcated into two branches.
Although this behavior is common to the radial mode,
it is apparent from the left panel of this figure that
the radial mode does not belong to the same family.
In the right panels, the $\ell$-dependences of the growth rate (upper panel)
and the luminosity (lower panel), at which the mode with a given $\ell$
becomes unstable for the first time, are shown.
The growth rate increases at first and then decreases as $\ell$ becomes larger.
The peak occurs at $\ell=5-11$, depending on the neutrino luminosity.
On the other hand, the luminosity, at which the convective instability appears,
decreases with $\ell $ at first until $\ell=6$ and then it increases.
These peaks are rather broad, and a couple of modes become unstable 
almost at the same luminosity with similar growth rates. 

As shown in the previous paper \citep{yam05,yam06},
there appears a heating region in the accretion flow
for the luminosity larger than a certain value.
It is easy to see that the heating region, if exists,
has a negative entropy gradient.
In such a region, the conventional criterion predicts
an existence of convection.
In our models with the mass accretion rate of $\dot{M}=1.0M_{\odot}/{\rm s}$, 
the critical luminosity for the existence of the heating region is
$3.7\times 10^{52}{\rm ergs/s}$.
It should be emphasized here that the actual convective instability occurred
at larger neutrino luminosity, $\sim 7\times 10^{52}{\rm ergs/s}$.
This confirms the recent findings by \citet{fog06a} that the advection
stabilizes the convective modes.

Introducing the parameter $\chi$, which is roughly the ratio of the advection
time scale and the growth time scale of convective instability,
\citet{fog06a} gave the additional criterion, $\chi >3$,
for the convective instability in the advecting matter.
In our previous paper \citep{yam06}, we calculated the value of $\chi$
for the steady solution that is employed in the present paper,
and found that the value of $\chi$ is always less than $3$.
If the criterion by \citet{fog06a} is applicable to our models,
we should have found no unstable convective mode. 
The apparent discrepancy is traced back to the difference of
the unperturbed models adopted for linear analysis.
\citet{fog06a} assumed simplified formulae for the neutrino heating and
equation of state while we employed more realistic ones and, as a result,
the post-shock flows are much complicated (see Fig.~2 of \citet{yam06}).
Another important difference is that we considered the whole region
between the standing shock and the neutrino sphere, whereas \citet{fog06a}
treated only the heating region, imposing a boundary condition
at the gain radius.
Our models suggest that the critical value of $\chi$ for the convective
instability may be as small as $0.356$.
Incidentally, \citet{fog06a} estimated that the growth rate is largest
at $\ell=6$, which is in rough agreement with our results although our
calculations showed that the value of $\ell$ depends on the neutrino luminosity.

\section{Dependence on the Assumptions}\label{dep}
 
In this section we discuss possible changes of the results obtained so far 
when different inner boundary conditions and perturbations of neutrino
luminosity are employed. 

\begin{figure}
\plotone{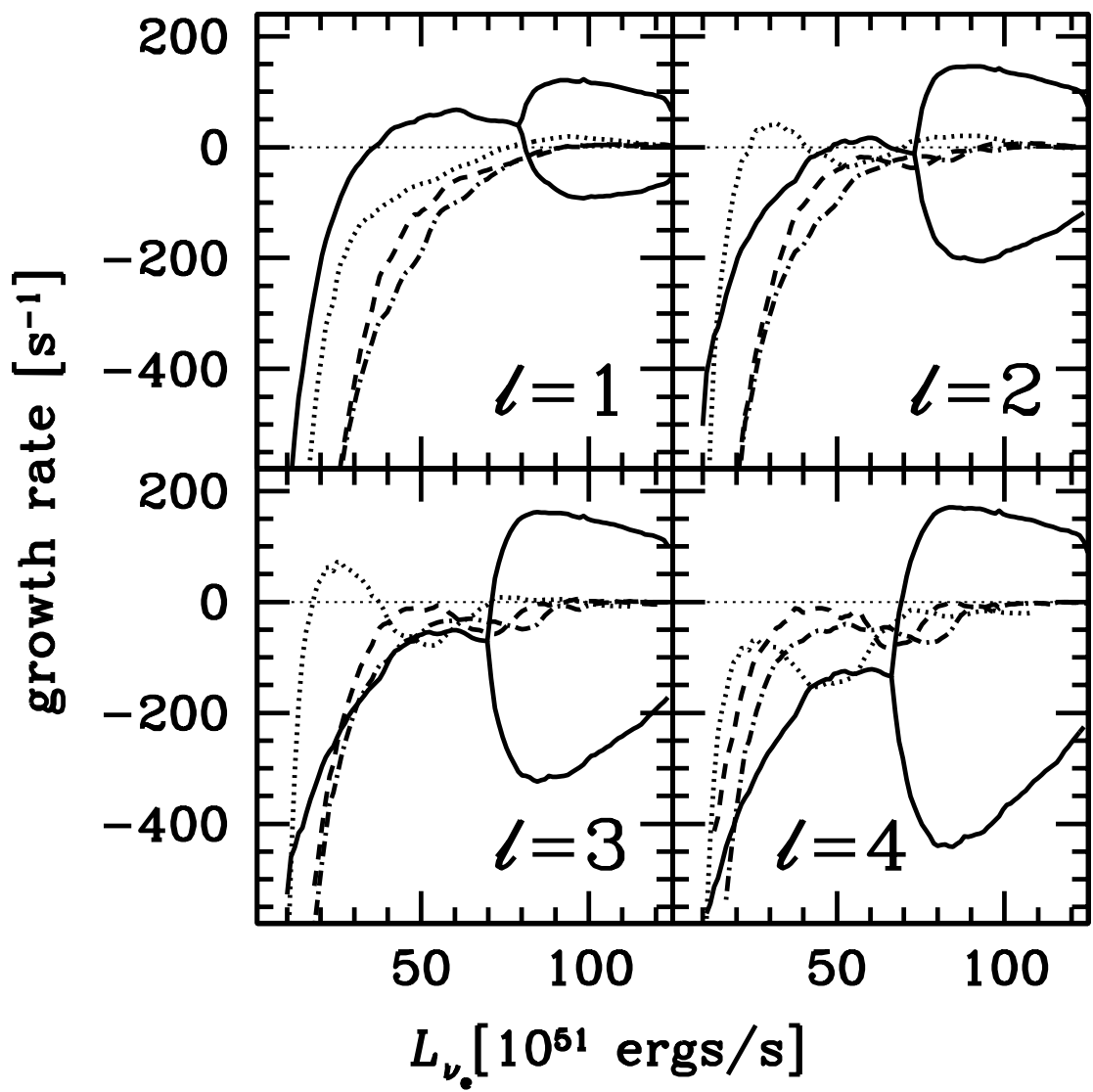}
\caption{The growth rates, ${\rm Re}(\omega)$,
for the non-radial modes of $\ell=1,2,3,4$
as a function of neutrino luminosity.
The solid curves represent the fundamental modes while the 
dotted, dashed and dash-dotted curves correspond to the first, second
and third overtones, respectively.
\label{fig2}}
\end{figure}

\begin{figure}
\plotone{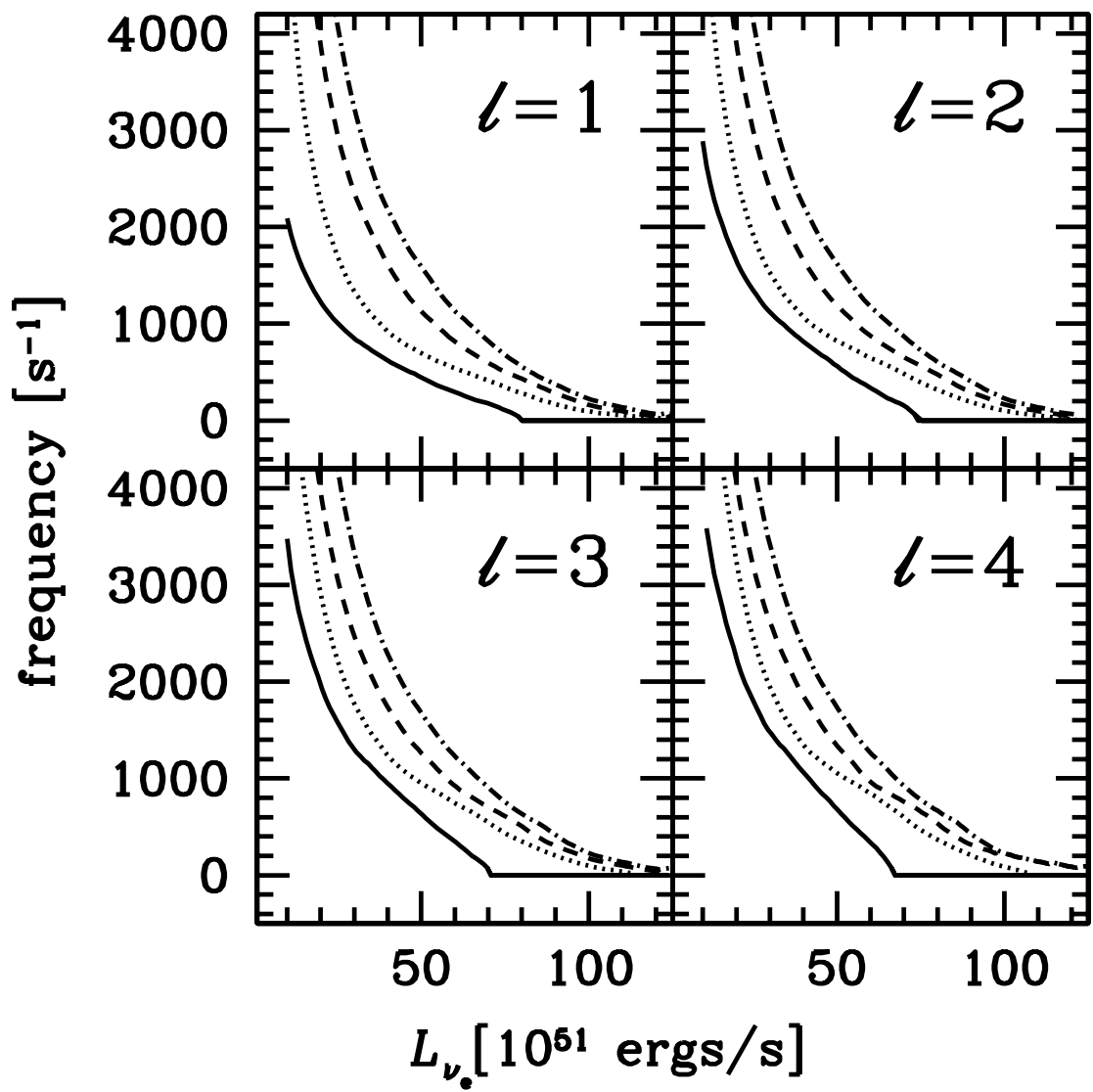}
\caption{The oscillation frequencies, ${\rm Im}(\omega)$,
the non-radial modes of $\ell=1,2,3,4$
as a function of neutrino luminosity.
The solid curves represent the fundamental modes while the 
dotted, dashed and dash-dotted curves correspond to the first, second
and third overtones, respectively.
\label{fig3}}
\end{figure}

\begin{figure}
\plotone{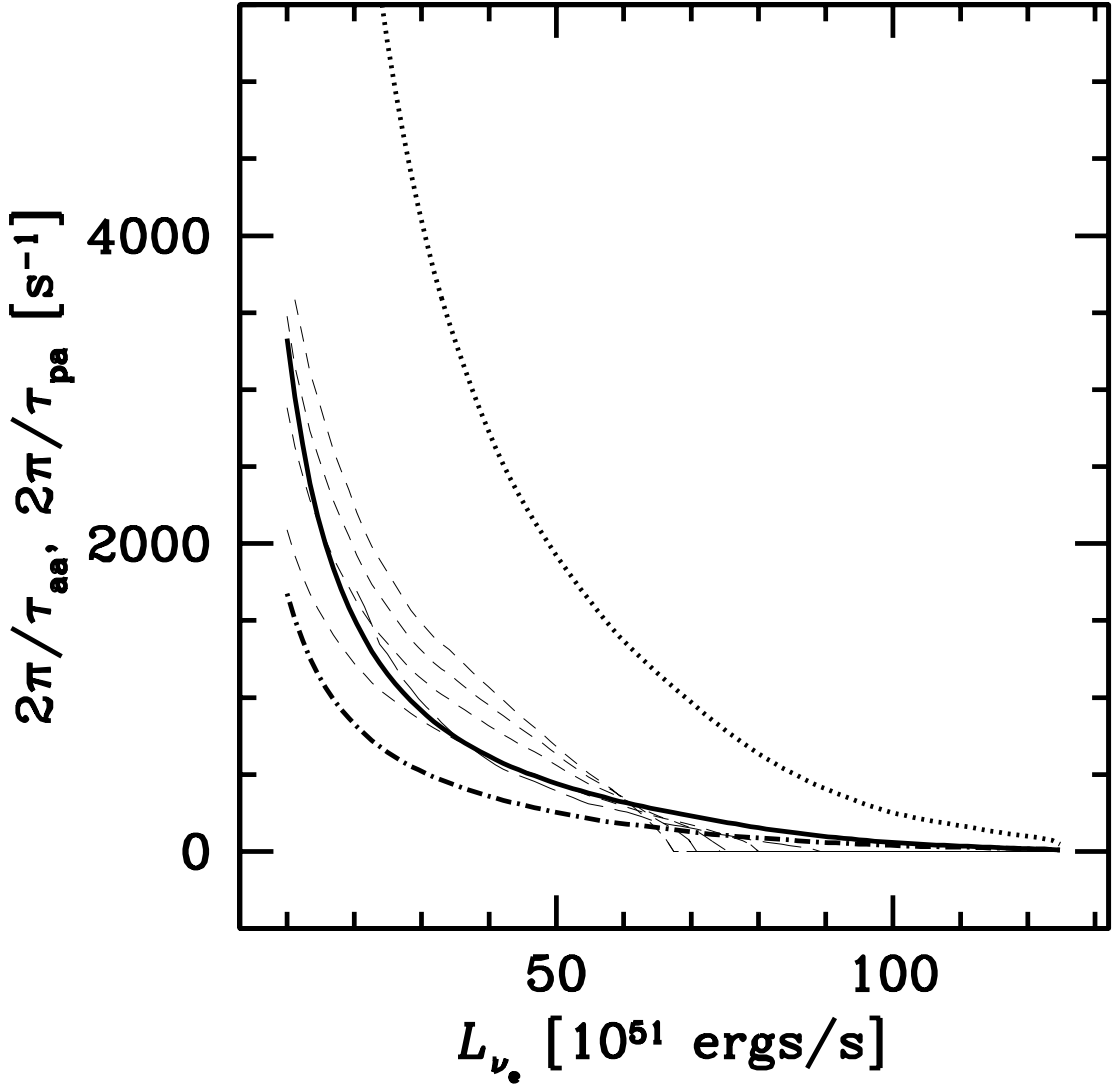}
\caption{The characteristic frequencies of the advection-acoustic cycle,
$2\pi/\tau_{\rm aa}$, and purely acoustic cycle, $2\pi/\tau_{\rm pa}$,
together with the oscillation frequencies for some modes.
The solid curve shows $2\pi/\tau_{\rm aa}$ while the dotted curve gives
$2\pi/\tau_{\rm pa}$.
The thin dashed curves represent the oscillation frequencies for the radial
(long dashed curve) and non-radial fundamental modes of $\ell=1,2,3,4$
from bottom to top for low neutrino luminosities. The dash-dotted
curve presents the characteristic frequency for the round trip
along the shock front for the sound wave.
\label{fig4}}
\end{figure}

\begin{figure}
\plotone{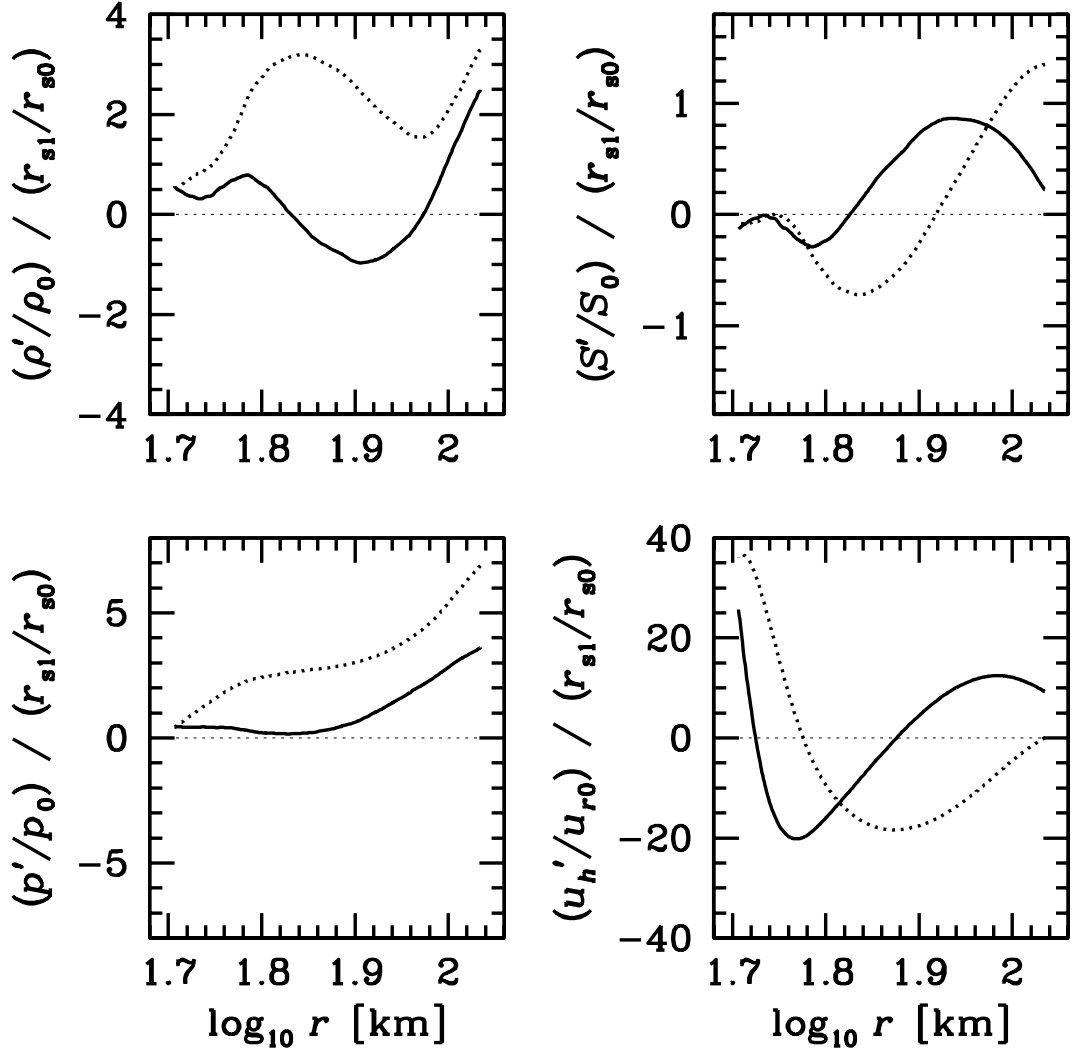}
\caption{The eigen functions of the non-radial unstable mode with $\ell=1$
for $L_{\nu_{\rm e}}=6\cdot 10^{52}$ergs/s. The fundamental mode is shown.
The solid and dotted curves represent the real and imaginary parts,
respectively.
\label{fig5}}
\end{figure}

\begin{figure}
\plotone{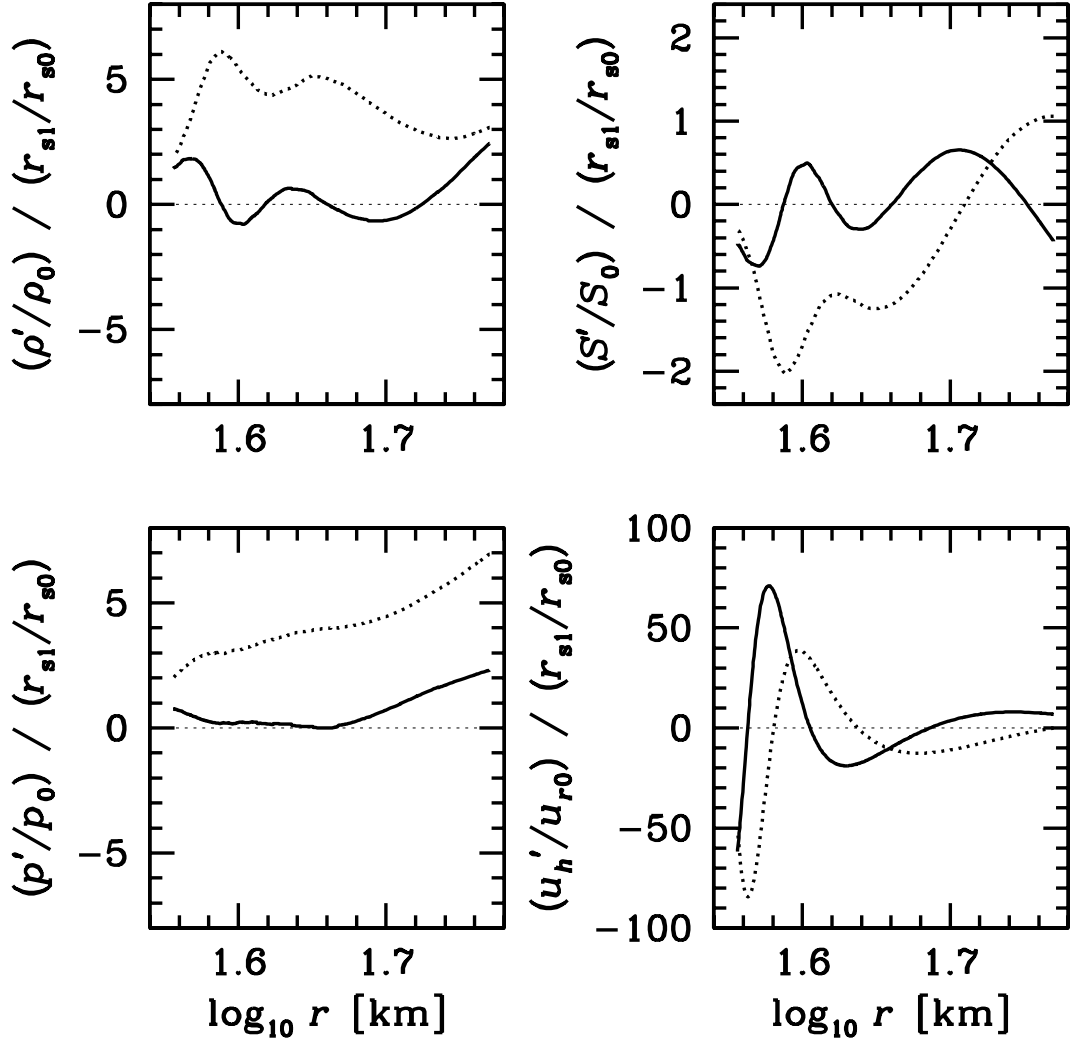}
\caption{The eigen functions of the non-radial unstable mode with $\ell=2$
for $L_{\nu_{\rm e}}=3\cdot 10^{52}$ergs/s. The first overtone is shown.
The solid and dotted curves represent the real and imaginary parts,
respectively.
\label{fig6}}
\end{figure}

\begin{figure}
\plottwo{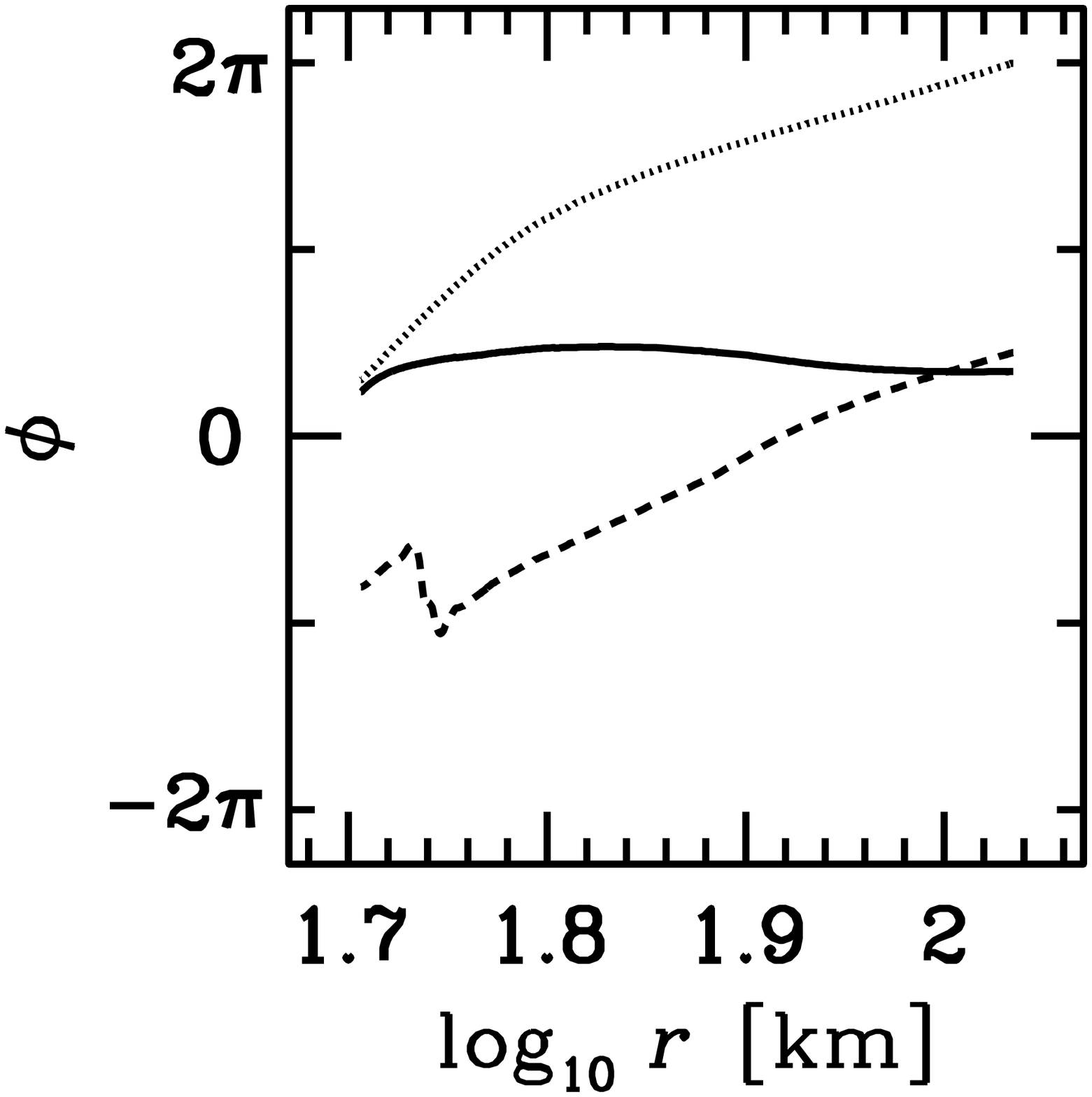}{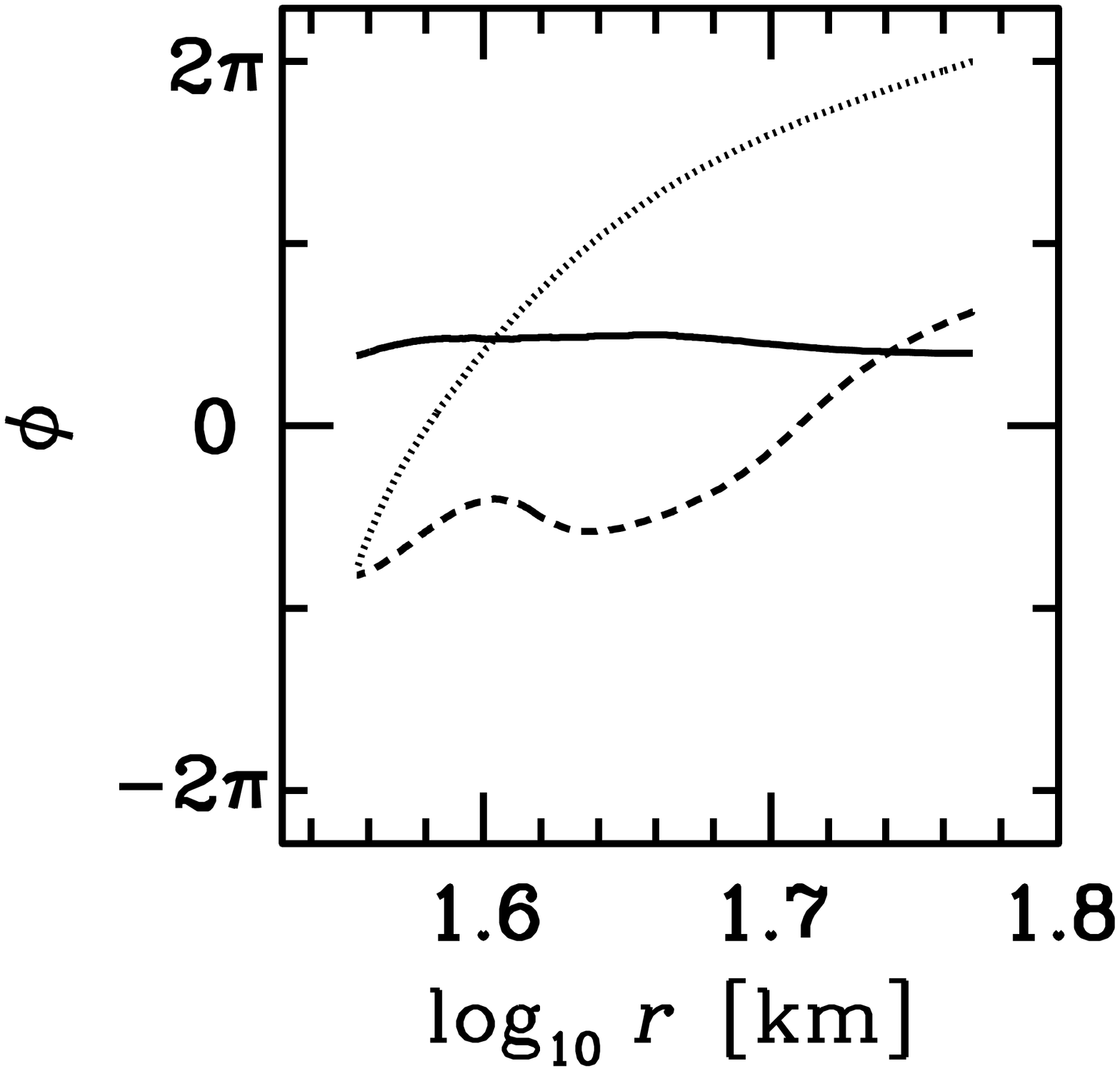}
\caption{{\it Left:} The radial profiles of the phases of various perturbed 
quantities for the fundamental mode of $\ell=1$ at 
$L_{\nu_{\rm e}}=6\cdot 10^{52}$ergs/s.
The solid, dotted and dashed curves represent the phases of
pressure, horizontal velocity and entropy perturbations, respectively.
{\it Right:} The same as in the left panel
but for the first overtone of $\ell=2$ at 
$L_{\nu_{\rm e}}=3\cdot 10^{52}$ergs/s.
\label{fig7}}
\end{figure}

\begin{figure}
\plottwo{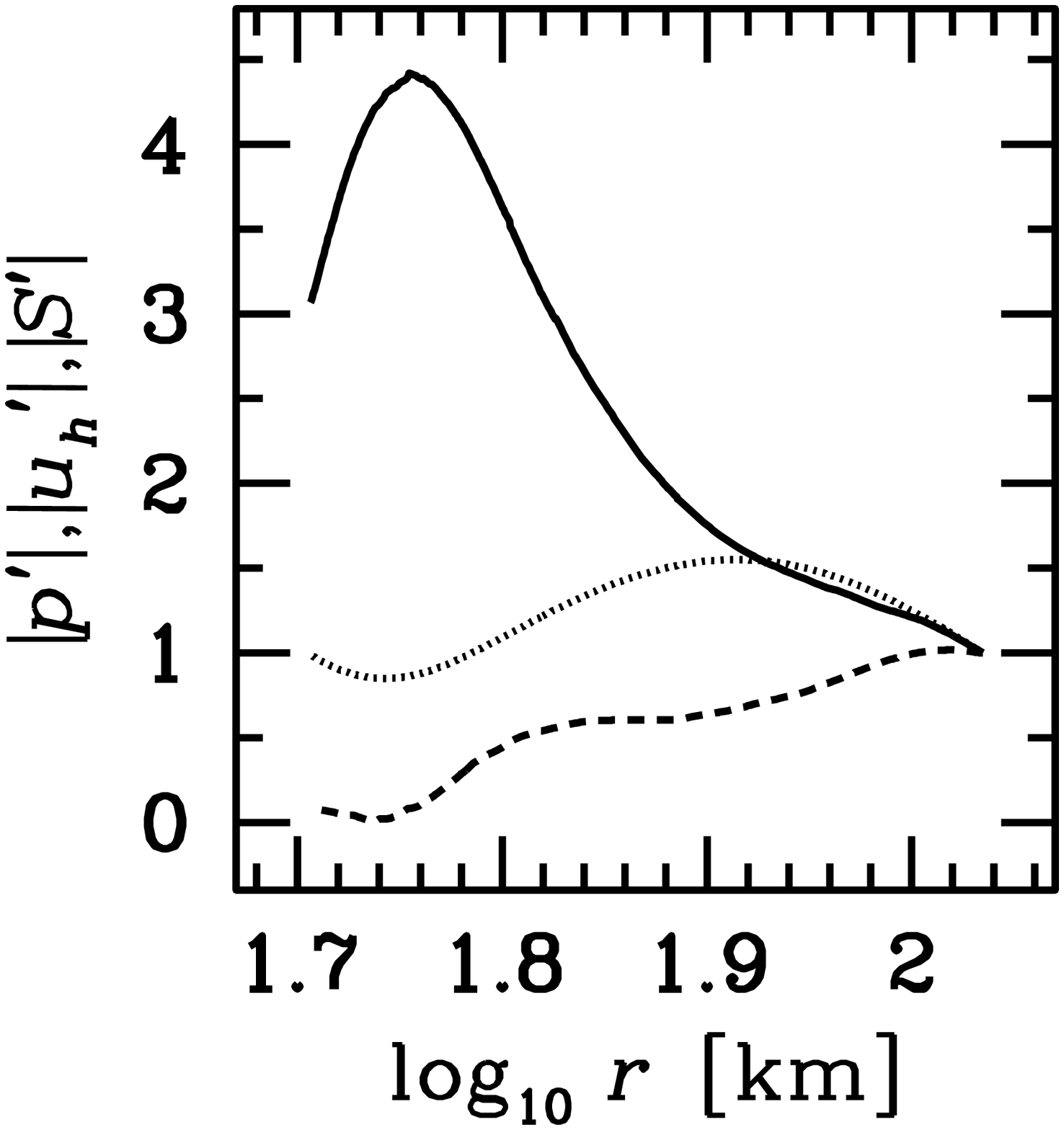}{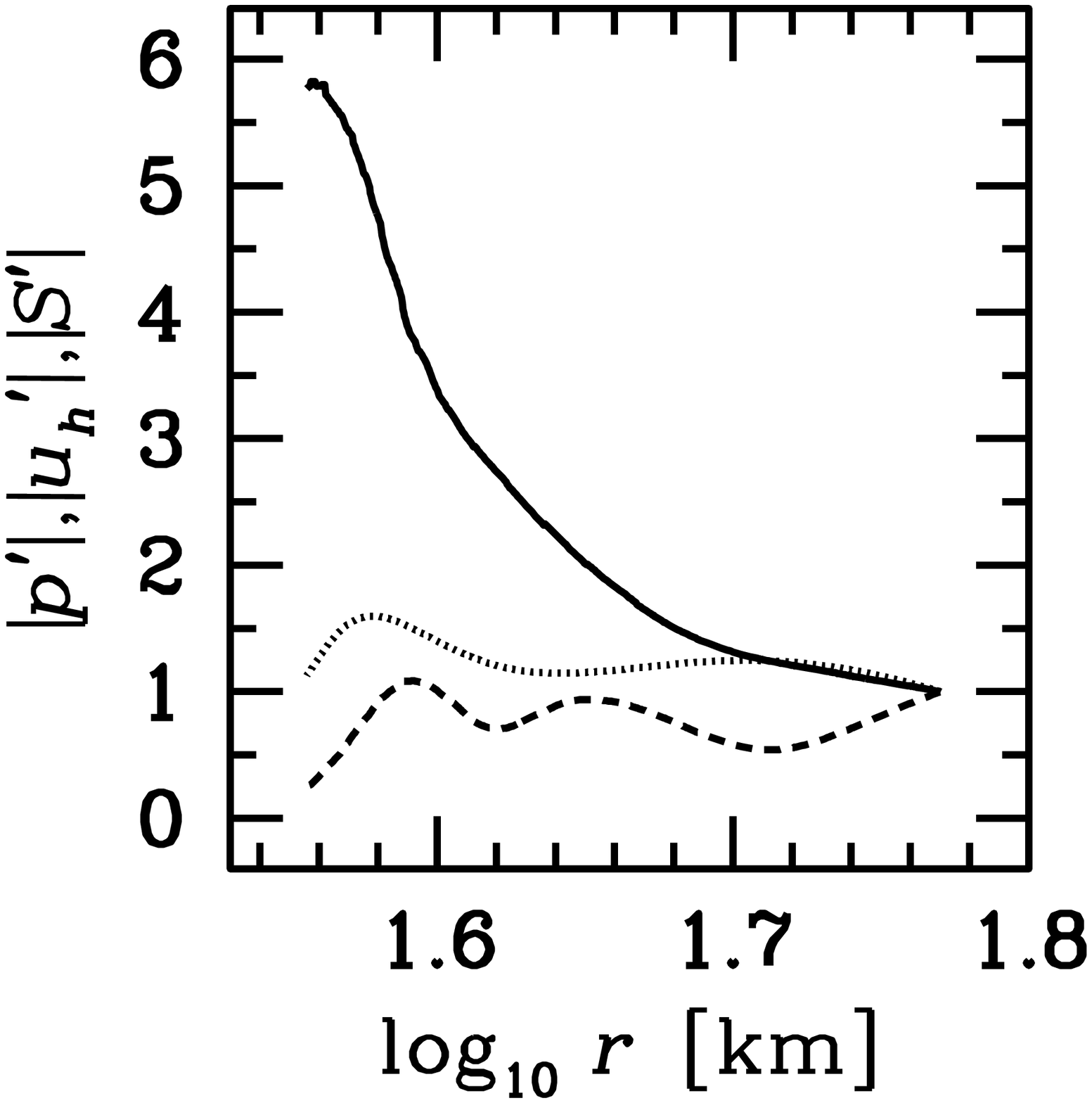}
\caption{{\it Left:} The radial profiles of the amplitudes of various perturbed
quantities for fundamental mode of $\ell=1$
at $L_{\nu_{\rm e}}=6\cdot 10^{52}$ergs/s.
The notation is the same as in Fig.~\ref{fig7}. 
The amplitudes are normalized at the shock front.
{\it Right:} The same as in the left panel
but for the first overtone of $\ell=2$
at $L_{\nu_{\rm e}}=3\cdot 10^{52}$ergs/s.
\label{fig8}}
\end{figure}

\begin{figure}
\plottwo{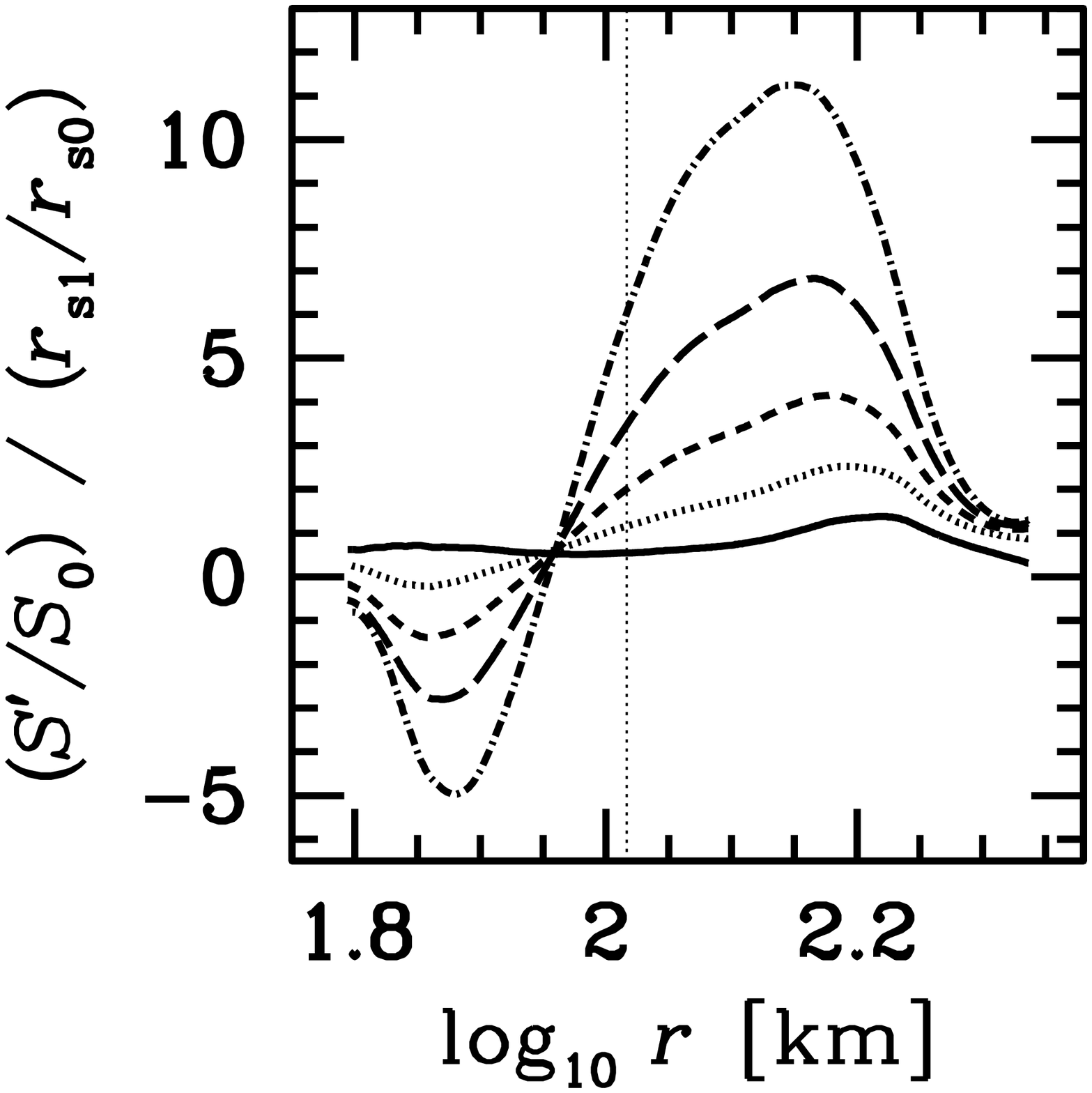}{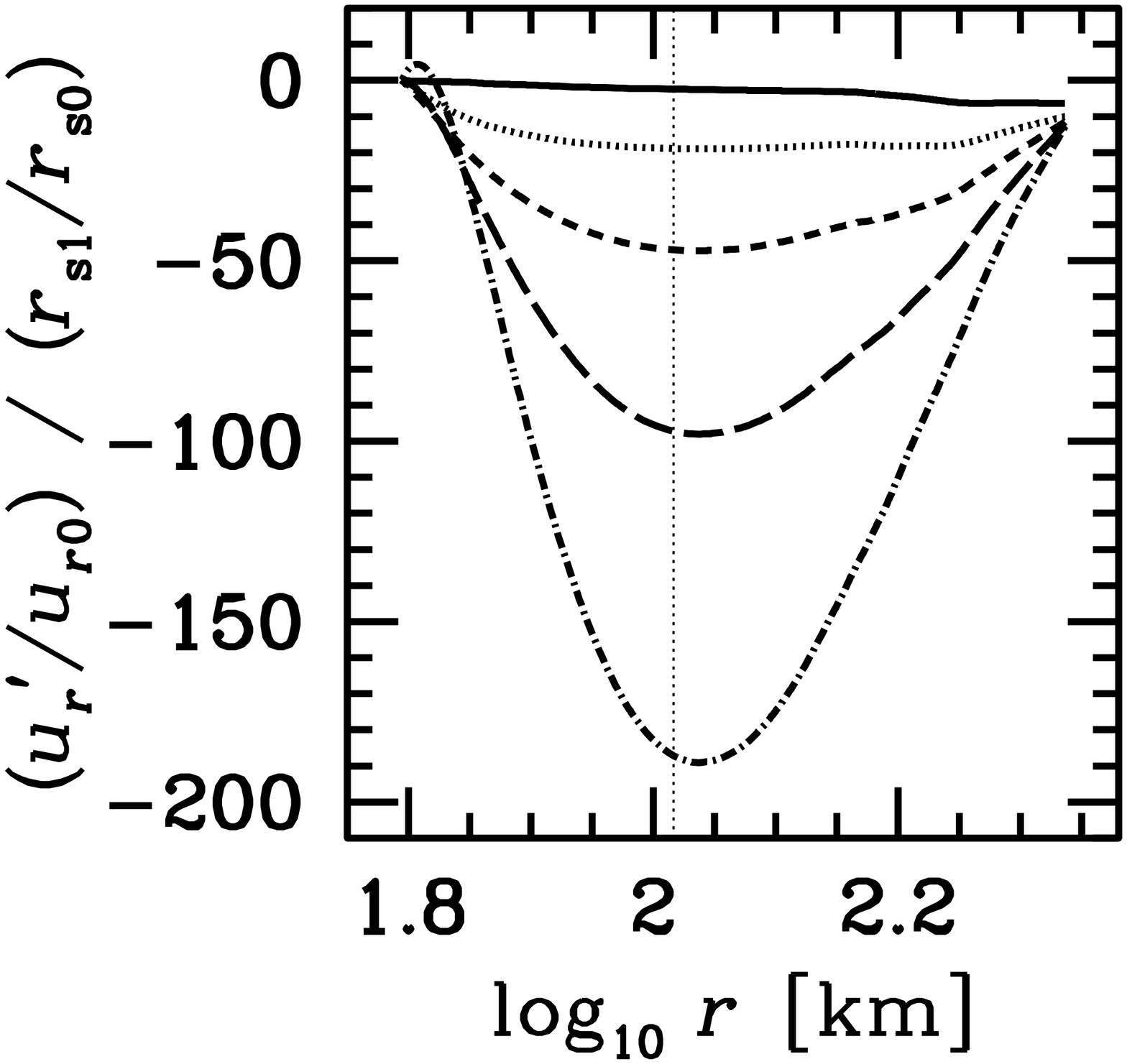}
\caption{{\it Left:} The eigen functions of the entropy perturbation
for non-oscillatory modes
at $L_{\nu_{\rm e}}=9\cdot 10^{52}$ergs/s.
The solid curve represents the radial model ($\ell=0$).
The dotted, dashed, long dashed and dash-dotted curves correspond to
the non-radial modes with $\ell=1,2,3,4$, respectively.
Thin dotted line indicates the location of the gain radius.
{\it Right:} The same as in the left panel
but for the perturbation of radial velocity.
\label{fig9}}
\end{figure}

\begin{figure}
\plottwo{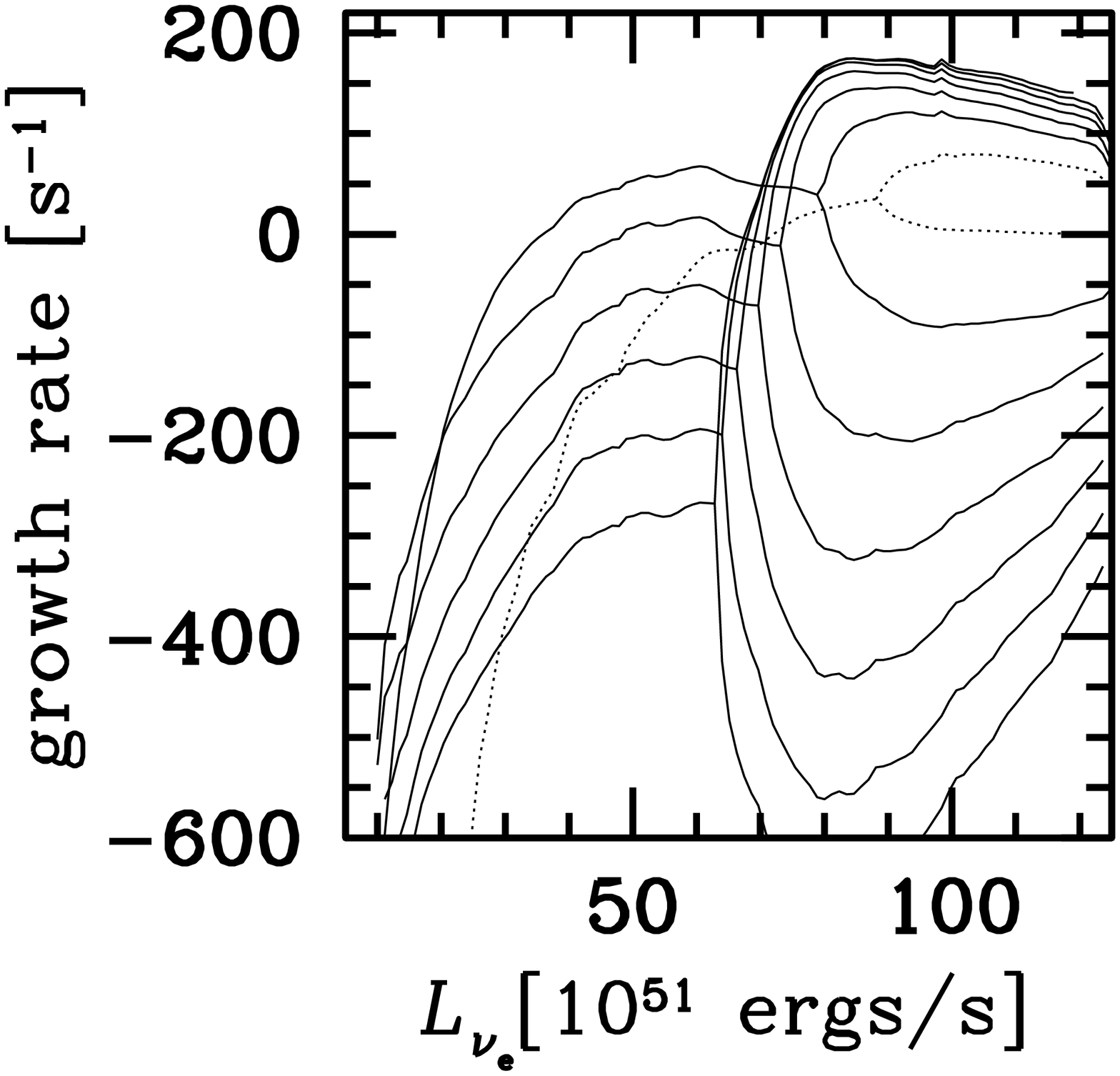}{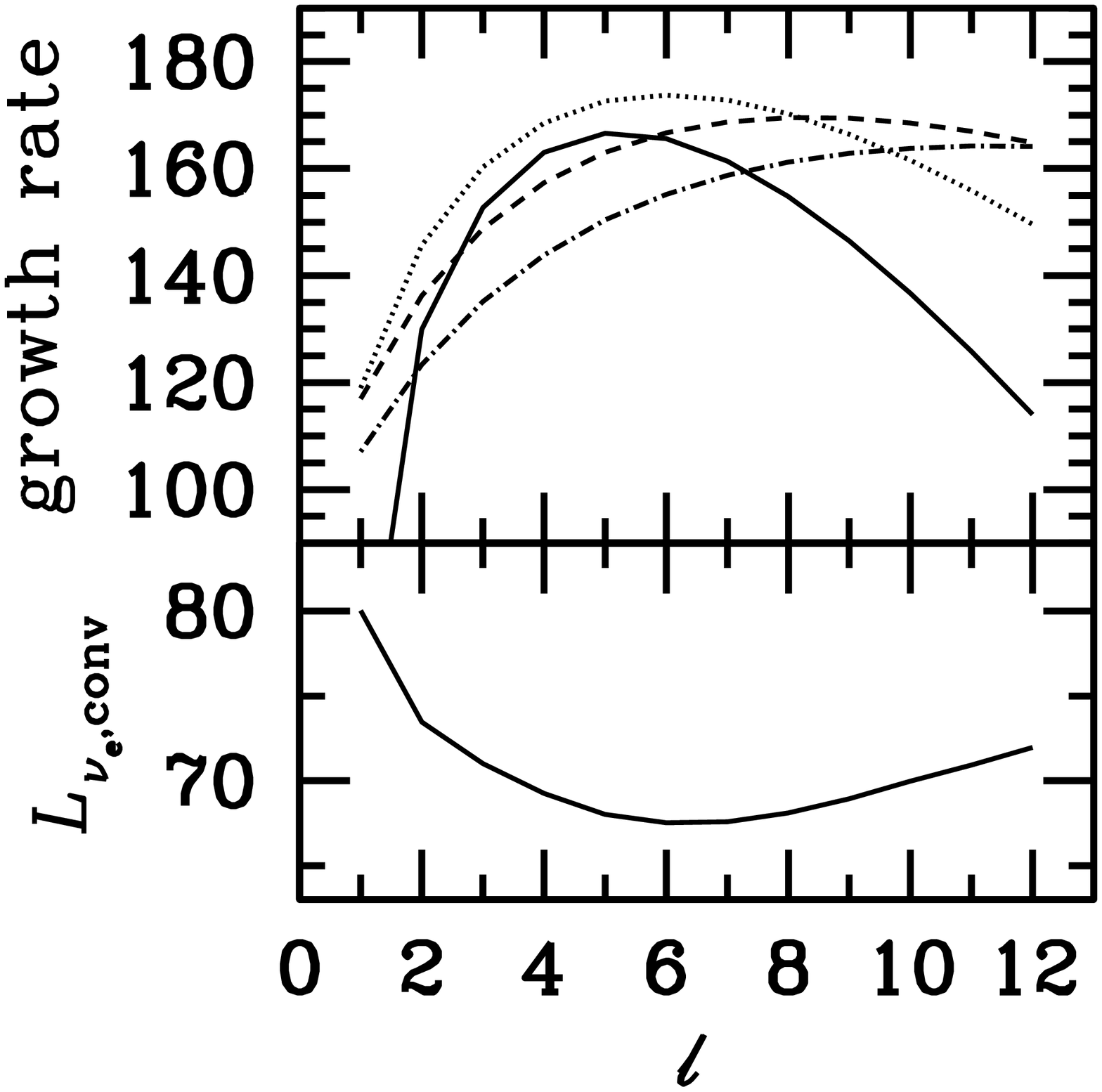}
\caption{{\it Left:} The growth rates, ${\rm Re}(\omega)$,
of the fundamental modes with different $\ell$'s.
The solid curves represent the non-radial modes with $\ell=1-6$ 
from top to bottom at $L_{\nu_{\rm e}}=5\cdot 10^{52}$ergs/s.
The dotted curve shows the radial mode ($\ell=0$). 
{\it Top right:} The growth rates, ${\rm Re}(\omega)$,
of the unstable non-oscillatory modes
(the upper branches of the bifurcated modes in the left panel)
as a function of $\ell$.
The solid, dotted, dashed and dash-dotted curve correspond to the results for 
$L_{\nu_{\rm e}}=8,9,10,11\cdot 10^{52}$ergs/s, respectively.
{\it Bottom right:} The neutino luminosity in unit of $10^{51}$ergs/s,
at which the convective mode with a given $\ell$ 
becomes unstable for the first time.
\label{fig10}}
\end{figure}

First of all, it should be mentioned that we imposed the solid boundary condition 
given by Eq.~(\ref{eq27}) at the neutrino sphere
where the inflow velocity does not vanish in the unperturbed flow.
Although this condition should have been imposed on the neutron star surface, the above treatment 
was necessary, however, because the accretion flow is not steady near the neutron star surface 
in reality and the structure of the boundary layer cannot be described appropriately
by the steady solution. As pointed out by \citet{fog05}, such a solid 
boundary condition at a place with a non-vanishing inflow velocity produces 
additional pressure waves there,  which typically grow on the advection time scale
and affect the growth rate of the vortex-acoustic cycle if the advection time is comparable to
the growth time as in the present cases. We can estimate this boundary effect by 
shifting the location of the inner boundary inwards \citep{fog06b},
since the advection time will then be increased and become sufficiently longer
than the growth time scale of the instability.
It is noted, however, that our original formulae
for the neutrino heating cannot be employed inside the neutrino sphere,
since the geometrical factor $f$,
cannot be employed inside the neutrino sphere, since the geometrical factor $f$,
which is given by Eq.~(\ref{eqa16}), then becomes imaginary. 
In order to avoid this difficulty,
we modified the geometrical factor as follows,
\begin{equation}
f=\frac{1}{4}\left(\frac{r_{\nu}}{r}\right)^2, 
\label{eq34}
\end{equation}
which corresponds to the asymptotic expression
for large $r$ of the original formula and
is correct if neutrinos are propagating only radially.
With this modification, we can extend the steady solution
inwards continuously and smoothly 
although it is a very rough approximation to the actual situation
inside the neutrino sphere.
It should be mentioned, therefore,
that the following models are of experimental nature and 
are not meant to be realistic. It is emphasized, however,
that the treatment ensures 
no artificial generation of sound waves occurs at the neutrino sphere.
Apart from this minor modification,
we solved the same equations as given in \S\ref{form} 
as well as in the previous paper \citep{yam06},
imposing the same condition, Eq.~(\ref{eq27}), at the shifted inner boundary.
We can confirm that the results were little affected
by the choice of the geometrical factor
(compare the solid curve in the upper left pannel of Fig.~\ref{fig2}
and that in Fig.~\ref{fig16}).

\begin{figure}
\plotone{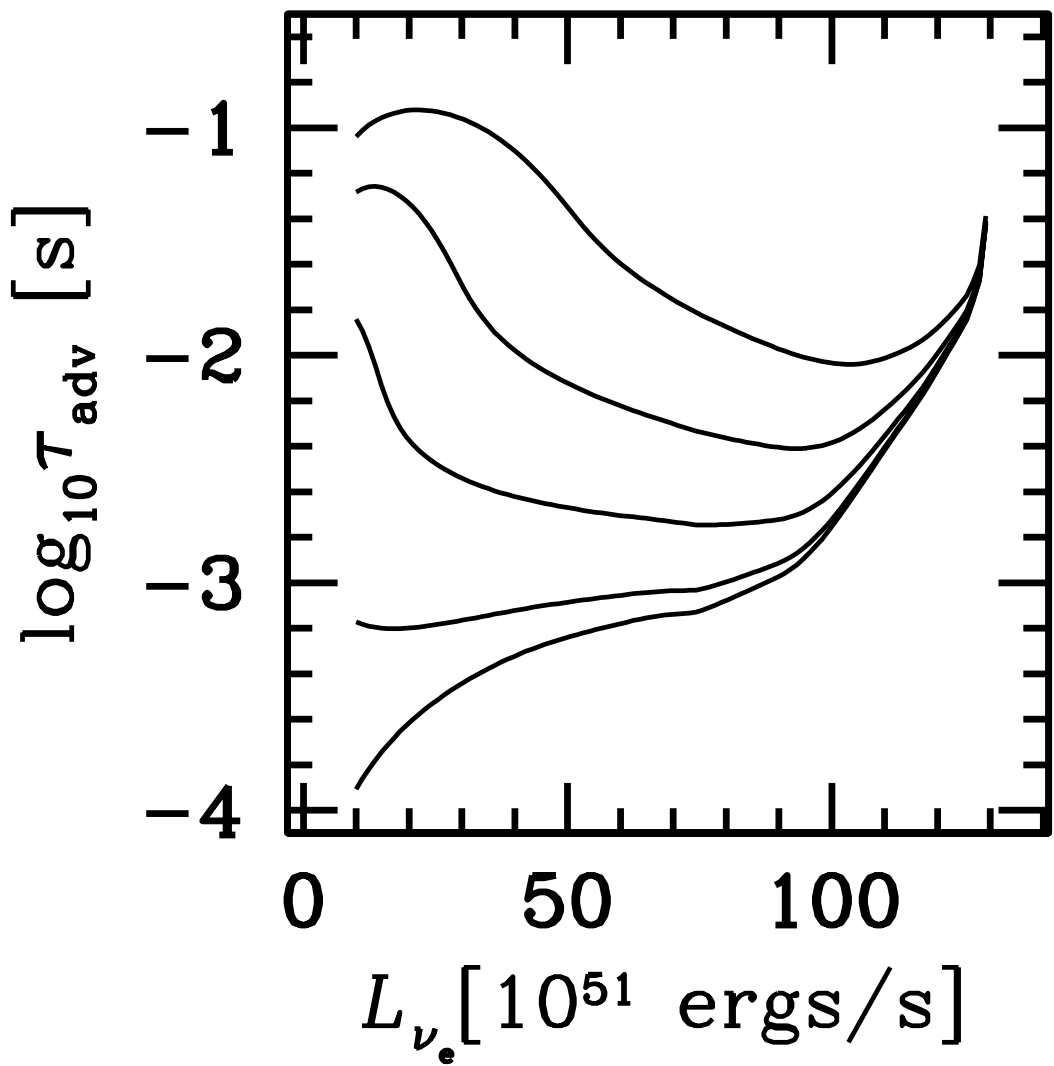}
\caption{The advection time, ${\tau}_{\rm adv}$,  
in the region between  shock and inner boundary.
The radius of the inner boundary is altered.
Each curve denotes the value for the case that the inner boundary
is placed at $1.0 r_{\nu}$, $0.9 r_{\nu}$, $0.8 r_{\nu}$, $0.7 r_{\nu}$,
$0.6 r_{\nu}$ from bottom to top, respectively.
\label{fig15}}
\end{figure}

\begin{figure}
\plotone{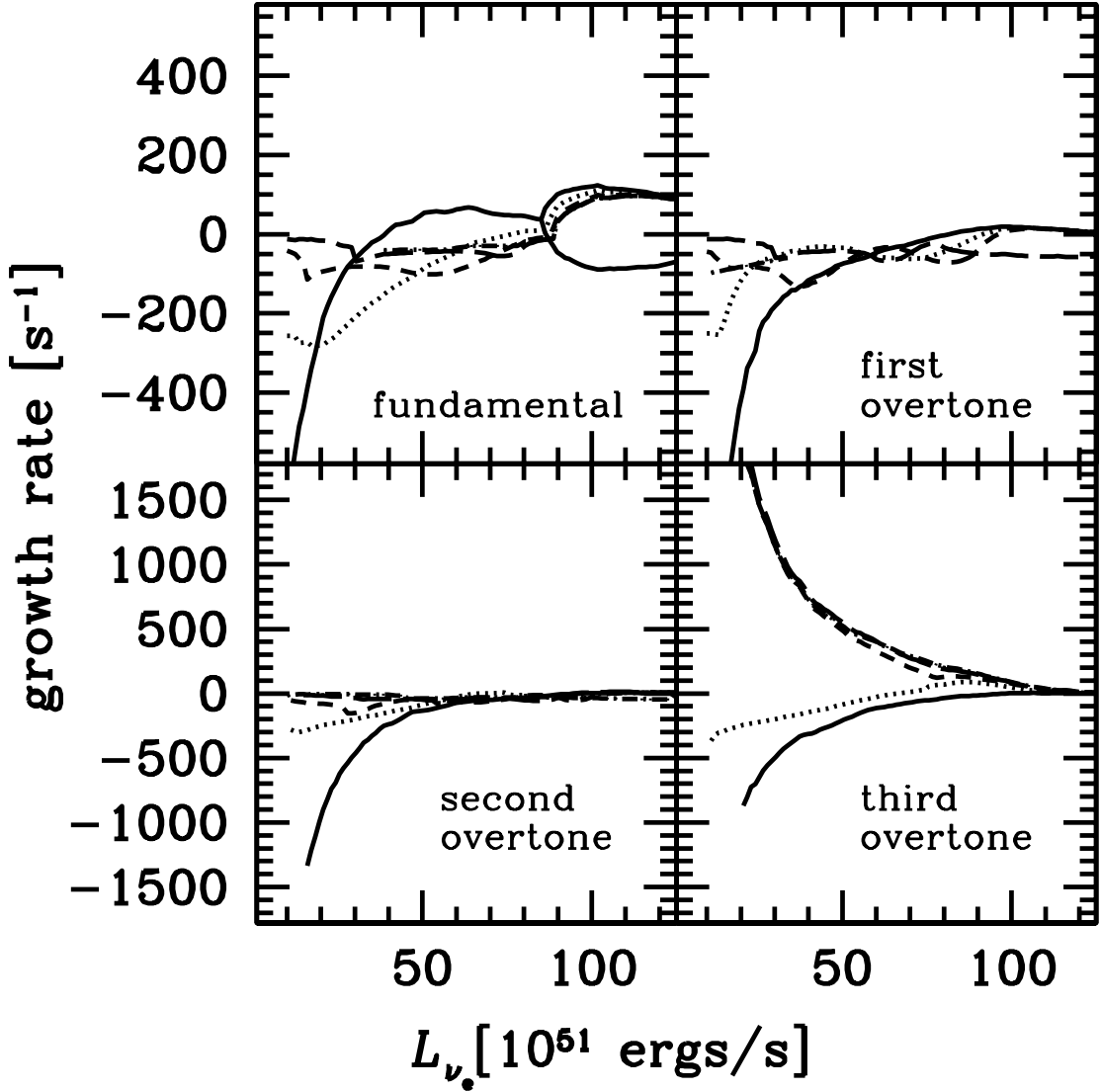}
\caption{The growth rates, ${\rm Re}(\omega)$,
of the non-radial modes with $\ell=1$
as a function of neutrino luminosity for different inner boundary radii. 
The fundamental mode (upper left panel),
first overtone (lower left panel), second overtone (upper right panel),
and third overtone (lower right panel) are shown.
The solid, dotted dashed, long dashed and dash-dotted curves
correspond to the result for the inner boundary radius of $1.0 r_{\nu}$, 
$0.9 r_{\nu}$, $0.8 r_{\nu}$, $0.7 r_{\nu}$, and $0.6 r_{\nu}$,
respectively. Notice that the vertical scales are different
between the upper and lower panels.
\label{fig16}}
\end{figure}

\begin{figure}
\plotone{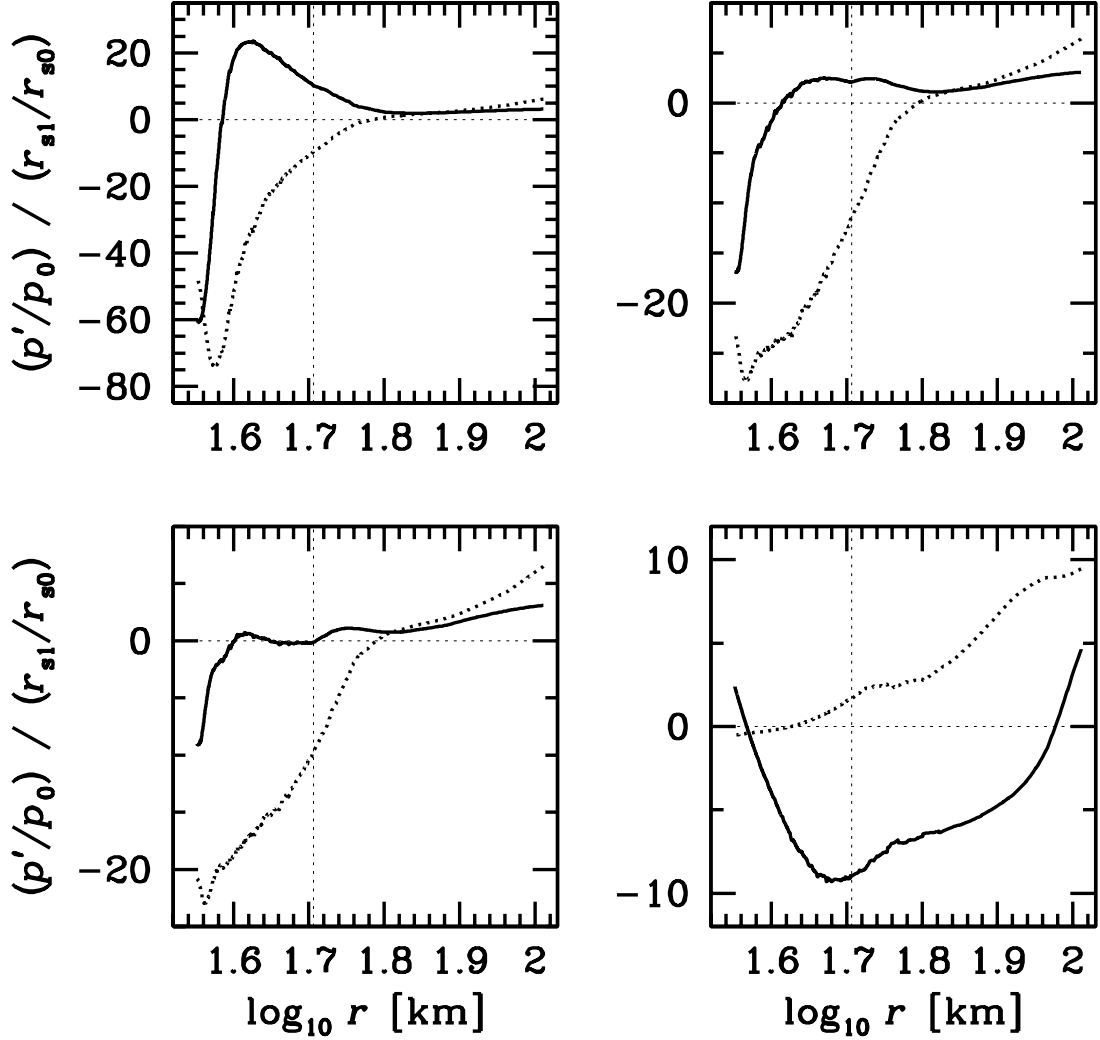}
\caption{The eigen functions of the non-radial
unstable mode with $\ell=1$
for $L_{\nu_{\rm e}}=6\cdot 10^{52}$ergs/s. 
The inner boundary is set at $r=0.7 r_{\nu}$.
The solid and dotted curves represent the real and imaginary parts,
respectively. The fundamental mode (upper left panel),
first overtone (upper right panel),
second overtone (lower left panel),
and third overtone (lower right panel) are shown.
The vertical thin dotted lines represent the location of the neutrino sphere.
\label{fig17}}
\end{figure}

\begin{figure}
\plotone{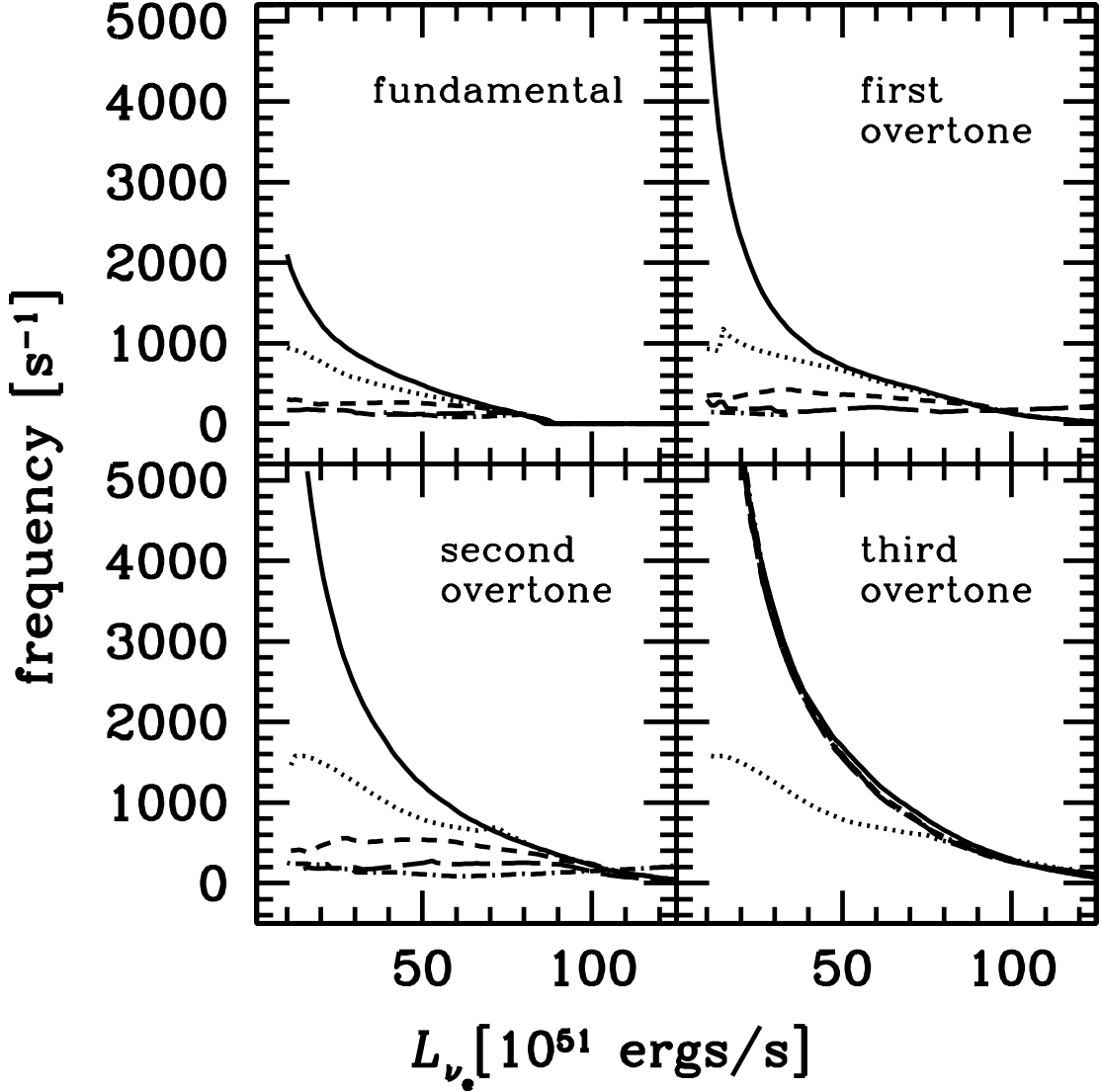}
\caption{The oscillation frequencies, ${\rm Im}(\omega)$,
of the non-radial modes with $\ell=1$
as a function of neutrino luminosity for different inner boundary radii. 
The fundamental mode (upper left panel),
first overtone (upper right panel), second overtone (lower left panel),
and third overtone (lower right panel) are shown.
The solid, dotted, dashed, long dashed and dash-dotted curves
correspond to the results for the inner boundary radius of $1.0 r_{\nu}$, 
$0.9 r_{\nu}$, $0.8 r_{\nu}$, $0.7 r_{\nu}$, and $0.6 r_{\nu}$,
respectively.
\label{fig18}}
\end{figure}

The results for the oscillatory modes are shown
in Figs.~\ref{fig15} - \ref{fig18}.
The advection time scales defined as
\begin{equation}
\tau_{\rm adv}\equiv-\int^{r_{\rm s}}_{r_{\rm in}}
\frac{1}{u_r}dr,
\label{eq35}
\end{equation}
are shown in Fig.~\ref{fig15}. Here $r_{\rm in}$ is
the radius of the inner boundary.
As expected, the time scale is drastically increased
as the location of the inner
boundary is shifted more and more inwards.
In Fig.~\ref{fig16}, the growth rates of the modes with $\ell=1$ are given.
We can see that the growth rates of the fundamental modes
are remarkably reduced as the radius of the inner boundary gets smaller.
We emphasize, however, that this does not imply that our previous results on 
the exsitence of the unstable modes were mere artefacts by the generation 
of sound waves at the inner boundary. Instead, what happened here is 
the change of characteristics of some eigen functions. 
In fact, as displayed in the upper left panel of Fig.~\ref{fig17},
the perturbation amplitude for the fundamental mode in the present case becomes substantial
only inside the neutrino sphere, which is qualitatively different from the correponding one
given in the lower left panel of Fig.~{\ref{fig5}}. 
As the radius of the inner boundary gets smaller,
the sound waves generated deep inside the gravitational well are
quickly damped as they propagate outwards.
As a result, the vortex-acoustic cycle is not excited by these pressure waves.
This is also the case for the first and second overtone modes as demonstrated
in the upper right and lower left panels of the figure. 

Sound waves are produced everywhere, though.
It should be noted that higher overtones put more weight 
on the sound waves generated at larger radii,
where they are not so severely damped. As a result, 
the vortex-acoustic cycle is excited not for the fundamental modes
but for some higher overtones in the present case. 
This is demonstrated in the lower right panel of Fig.~\ref{fig16},
where the third overtone is shown to be unstable. 
The eigen function for this mode is displayed
in the lower right panel of Fig.~\ref{fig17},
in which one can easily recognize its qualitative difference from the others.
In Fig.~\ref{fig18}, the oscillation frequencies of the modes with
$\ell=1$ are given for different inner boundary radii.
We can see that the oscillation frequencies of the unstable third overtone 
for $r_{\rm in} < 0.8r_{\nu}$ are almost unchanged and close to $\tau_{\rm aa}$
but different from $\tau_{\rm pa}$. (It is noted that the $\tau_{\rm pa}$ is
only slightly altered from those
shown in Fig.~{\ref{fig4}} by shifting the radius of the inner boundary 
because the dislocation is much smaller than the distance between the shock and
neutrino sphere and the sound velocity is larger inside
the neutrino sphere than outside.)
Hence the unstable third overtones appear to be excited
by the vortex-acoustic couplings
rather than by the purely acoustic cycles.
Moreover, since the growth times of these unstable modes are much shorter 
than the advection time, the instabilities should be hardly affected by
the production of sound waves at the inner boundary. 
They are indeed driven by the pressure perturbations generated
near the neutrino sphere. On the other hand, the comparison between the growth 
rates of these modes and that of the fundamental mode obtained in \S 4 suggests that 
the solid boundary condition imposed at the neutrino sphere tended to suppress the 
excitation of the vortex-acoustic cycles.
To summarize the effect of the shifted inner boundary on the oscillatory modes, 
the fundamental modes and some lower overtones are confined inside the neutrino sphere
whereas higher overtones take over the roles that the fundamental
and lower overtone modes have played in the previous models.

\begin{figure}
\plotone{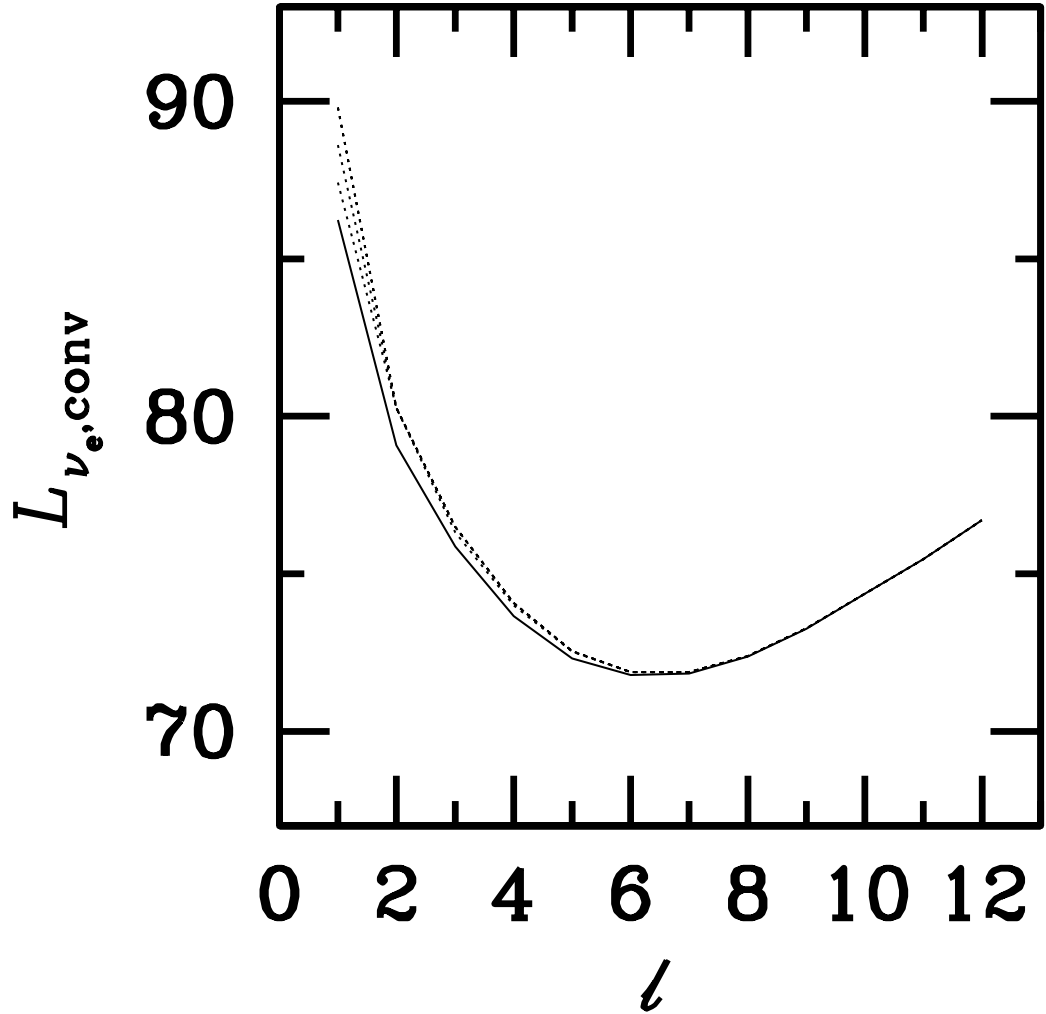}
\caption{The same as the right bottom panel of Fig.~\ref{fig10} but
for various values of the inner boundary radius.
The solid curve corresponds to the inner boundary radius of 
$r=1.0 r_{\nu}$. The dotted curves represent the results for the inner boundary radius of 
$r=0.9$, $0.8$, $0.7$, and $0.6 r_{\nu}$ from bottom to top, respectively.
\label{fig19}}
\end{figure}

On the other hand, the growth of non-oscillatory modes is only slightly
affected by the inward shift of the inner boundary (see Fig.~\ref{fig19}).
This is simply because the fluctuations for these modes grow
mainly at a distant region from the
neutrino sphere and the effects of over-generation of sound waves
at the inner boundary, if any, 
are negligible for the unstable region. Therefore, the quantitative,
but not qualitative, discrepancy between 
the results of \citet{fog06a} and ours on the stability criterion
for convection should be
attributed to the difference in the treatment of
the excitation of pressure perturbations 
between the gain radius and the neutrino sphere.
The more complex background model we employed here may 
have also contributed somewhat to the difference.

We have observed that the solid boundary condition imposed at the neutrino sphere
tends to stabilize the oscilatory modes and has essentially no influence
on the non-oscilatory modes.
The validity of the solid boundary condition itself is a remaining problem, though.
This is a difficult issue, which could not be addressed based
on the steady-state approximation,
and the complete answer will be beyond the scope of this paper.
In order to estimate the effect of 
different boundary conditions, however, we here adopt another conceivable condition, that is, 
the free boundary condition given by 
\begin{equation}
\frac{du'_r}{dr} = 0
\label{eq:eq33}
\end{equation}
at the inner boundary.
This made the calculation of the linear perturbations
much more difficult, however.
In fact, we often came across the difficulty in obtaining
a sufficient convergence of iterative calculations. 
As a result, the eigen values that we present below have much larger numerical
errors systematically, which will be inferred from the artificial fluctuations
in Fig.~\ref{fig14}.
In spite of the problem, we still think that we can  obtain a result
that is qualitatively correct and will be helpful for the society. 

In Fig.~\ref{fig14}, we plotted the growth rates, ${\rm Re}(\omega)$,
for the $\ell=1$ modes as a function of the neutrino luminosity.
This should be compared with the upper left panel of Fig.~\ref{fig2}.
The qualitative behavior of the fundamental mode is similar.
However, the growth rate of unstable oscillation is a bit larger
and the neutrino luminosity, at which the mode becomes unstable,
is a little lower than in the model with the solid boundary condition.
This seems to imply that the free inner boundary condition tends to
enhance the instability, contrary to the naive expectation.
This is also reflected in the behaviors of the overtones.
In fact, while in the solid boundary case 
the overtone modes are almost stable up to the critical luminosity,
in the free boundary case they are 
as unstable as the fundamental modes for relatively small luminosities.
These results with the free boundary indicate that
the qualitative results of the existence of the growing modes are not altered
if the condition is significantly altered and suggest that
the similar conclusion
is expected if more reasonable condition, if it exists, is imposed.

Next we consider the variation of the perturbations of neutrino luminosity.
In the previous sections, we assumed that the neutrino temperature
(and luminosity, as a result) was not affected by the perturbations.
This is based on the idea that neutrino luminosity is dominated
by the contribution from the proto neutron star,
which is little affected by the fluctuations of accretion flows.
It is noted that the neutrino temperature characterizing its spectrum
does not coincide in general with the matter temperature at the neutrino sphere.
It is possible, however, that the neutrino emission from the accreted matter
is more important and the fluctuation of the neutrino luminosity is correlated
with the perturbation of matter temperature.
Here we consider this case by imposing the condition that
the variation of the neutrino temperature is 
proportional to that of the matter temperature, that is,
\begin{equation}
\left.\frac{T_{\nu 1}}{T_{\nu 0}}=\frac{T_1}{T_0}\right|_{r=r_{\nu}}.
\label{eq32}
\end{equation}
We assumed that the temperature fluctuations given by Eq.~(\ref{eq32}) affect
only the heating of the matter with the same polar and azimuthal angle
($\theta$, $\phi$).

\begin{figure}
\plotone{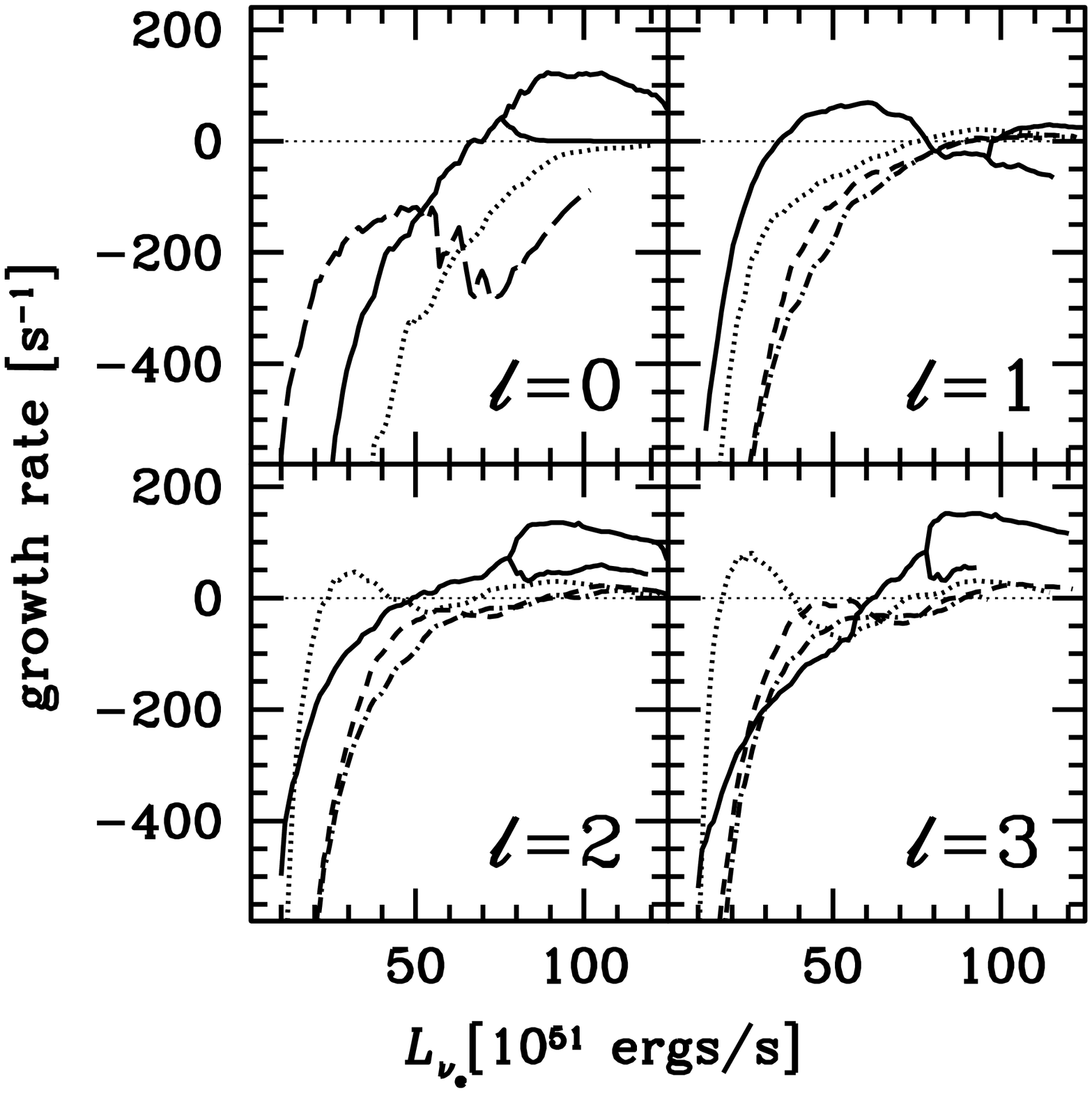}
\caption{The growth rates, ${\rm Re}(\omega)$,
for the radial, $\ell=0$, and non-radial modes, $\ell=1,2,3$,
as a function of neutrino luminosity. The perturbation of neutrino temperature 
is taken into account. In the upper left panel,
the solid curve represents the fundamental mode
and the dotted curve presents the first overtone.
The long dashed curve corresponds to the thermal mode.
In the other panels, the solid curves represent the fundamental modes
while the dotted, dashed and dash-dotted curves correspond to the first,
second and third overtones, respectively.
\label{fig11}}
\end{figure}

\begin{figure}
\plotone{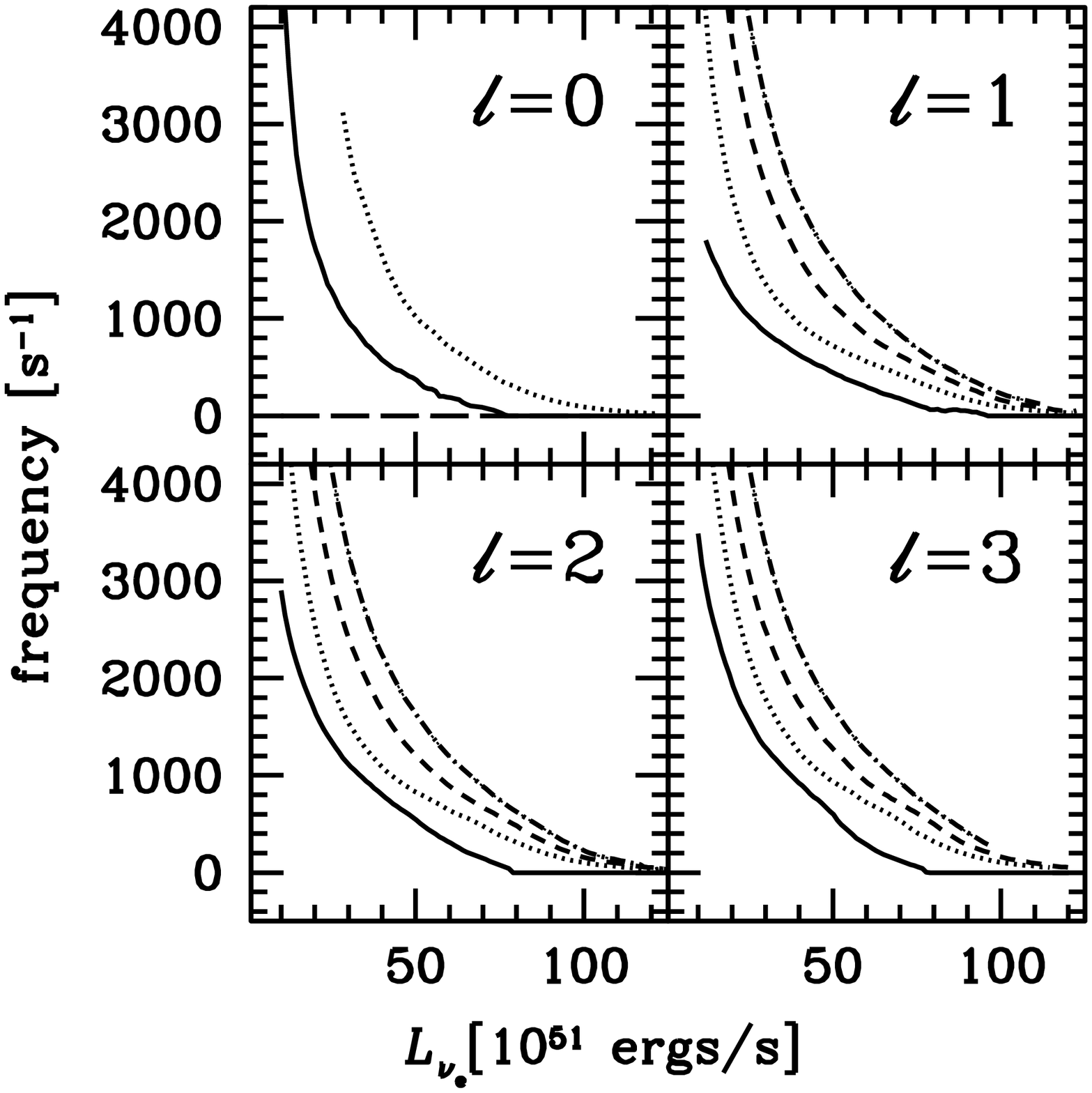}
\caption{The oscillation frequencies, ${\rm Im}(\omega)$,
for the radial, $\ell=0$, and non-radial modes, $\ell=1,2,3$,
as a function of neutrino luminosity. The perturbation of neutrino temperature 
is taken into account. In the upper left panel, the solid curve represents
the fundamental mode and the dotted curve presents the first overtone.
The long dashed curve corresponds to the thermal mode.
In the other panels, the solid curves represent the fundamental modes while the 
dotted, dashed and dash-dotted curves correspond to the first, second
and third overtones, respectively.
\label{fig12}}
\end{figure}

The results are shown in Figs.~\ref{fig11} and \ref{fig12}.
We show the growth rates and oscillation frequencies for the modes
with $\ell=0-3$.
Compared with the corresponding results shown in Figs.~\ref{fig2}
and \ref{fig3}, 
the qualitative feature is unchanged. The thermal mode exists
among the radial modes and is stable for all the neutrino luminosities
as in the previous case. The first overtones of $\ell=2,3$ modes become
unstable for relatively small luminosities.
Then the fundamental $\ell =1$ mode takes over.
They are all oscillatory. Again above $L_{\nu_{e}}\sim 7\times 10^{52}$ergs/s
the fundamental modes are bifurcated into two branches
and become non-oscillatory. 
Quantitatively, the growth rate is slightly modified with no clear tendency. 
The temporal variation of neutrino temperature sometimes helps
the growth of the perturbations but other times not.
It is determined by the integral of the fluctuations of heating over
the entire cycle of the oscillation of the neutrino temperature.
If the fluctuation of neutrino temperature and those of other quantities are
synchronized, the perturbation is enhanced and vice versa.
It should be also noted that the variation of the neutrino temperature is not
necessarily synchronized with the temperature of
the accreting matter at the neutrino sphere in reality.

We tried yet another case concerning the perturbation of neutrino emissions.
For both cases considered thus far,
the thermal mode was stable even at the critical luminosity, contraty to
the results of the previous paper \citep{yam05}
(see the long dashed lines for the $\ell=0$ modes
in Figs.~\ref{fig1} and \ref{fig11}).
Although there are several differences between the previous and current papers, 
the inner boundary condition is likely to be the most important.
Hence we took into account here the variation of the inner boundary radius
in such a way that the density-perturbation be zero at the inner boundary
for the perturbed flows.
This is exactly the same treatment we adopted
in the previous paper \citep{yam05}.
Note that this condition is difficult to impose for non-radial modes,
and we consider only the radial modes here.

\begin{figure}
\plottwo{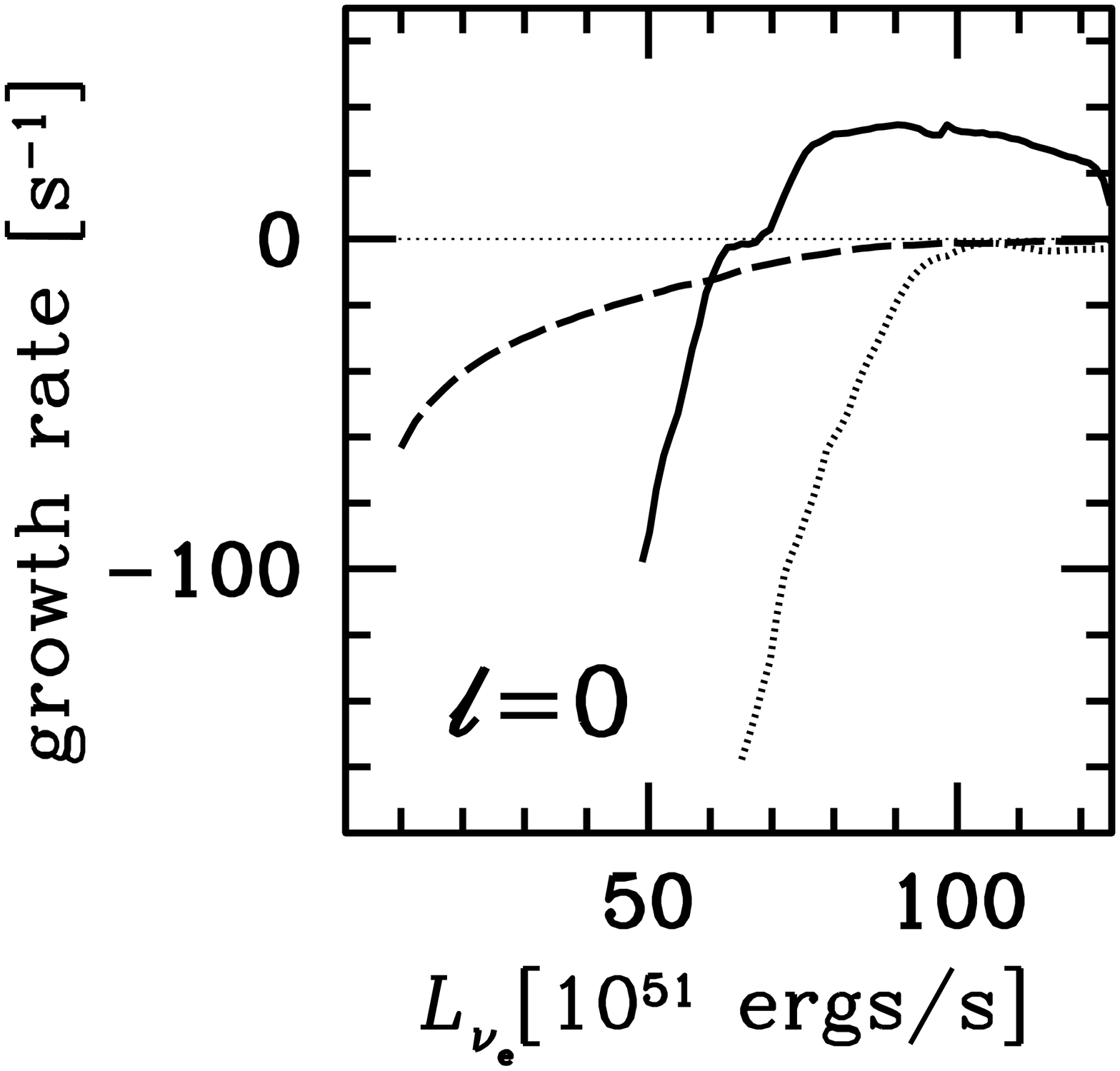}{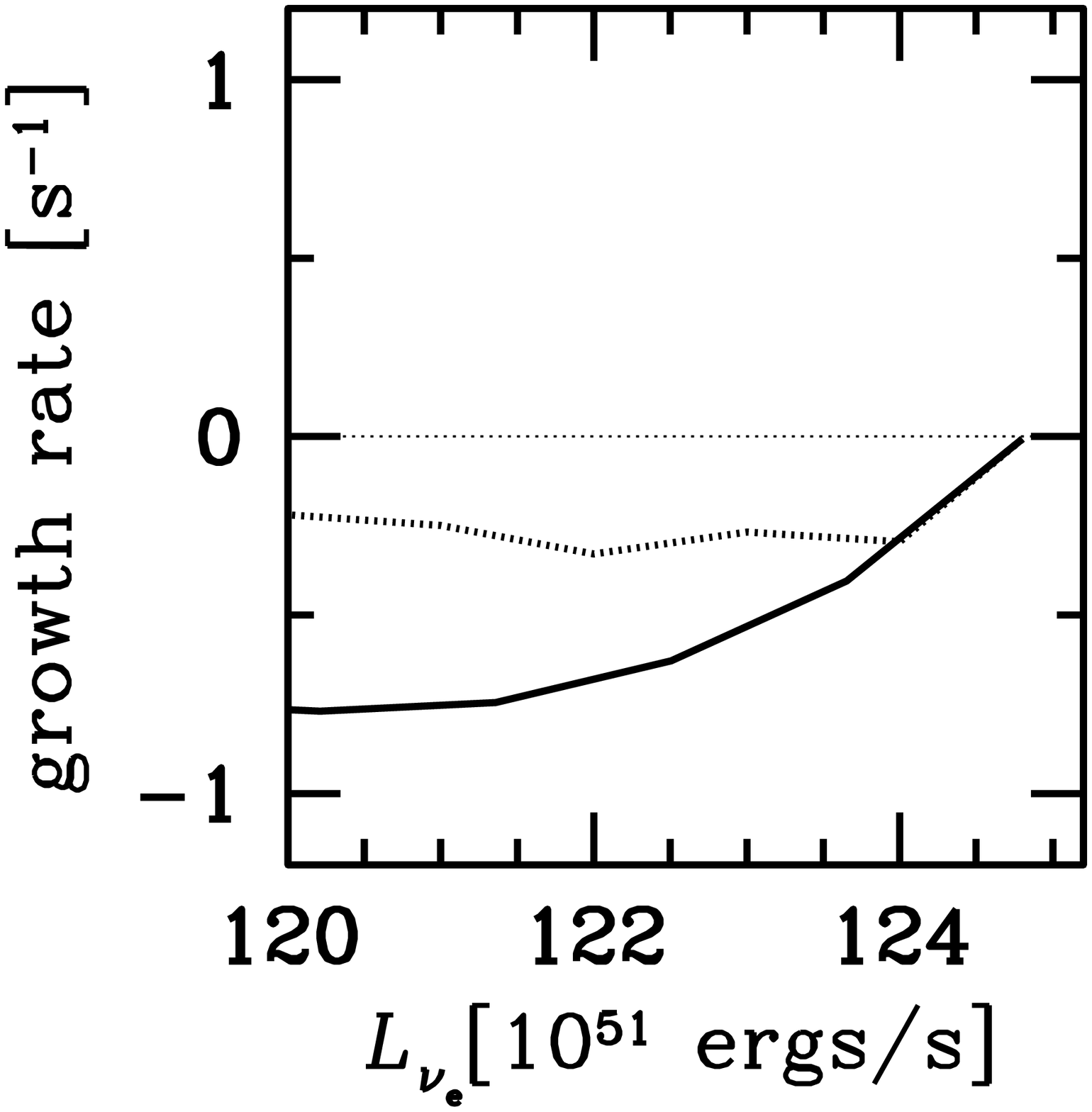}
\caption{{\it Left:} The growth rates, ${\rm Re}(\omega)$
for representative radial ($\ell=0$) modes as a function of neutrino luminosity.
The perturbations of neutrino temperature and neurtrino sphere are
taken into account.
The long dashed curve denotes the fundamental thermal mode.
The solid and dotted curves correspond to
the fundamental and first overtone oscillatory modes, respectively.
{\it Right:} The same as in the left panel, but
only those for the thermal modes
both for the inner solution (solid curve)
and the outer solution (dotted curve) are shown.
\label{fig13}}
\end{figure}

\begin{figure}
\plotone{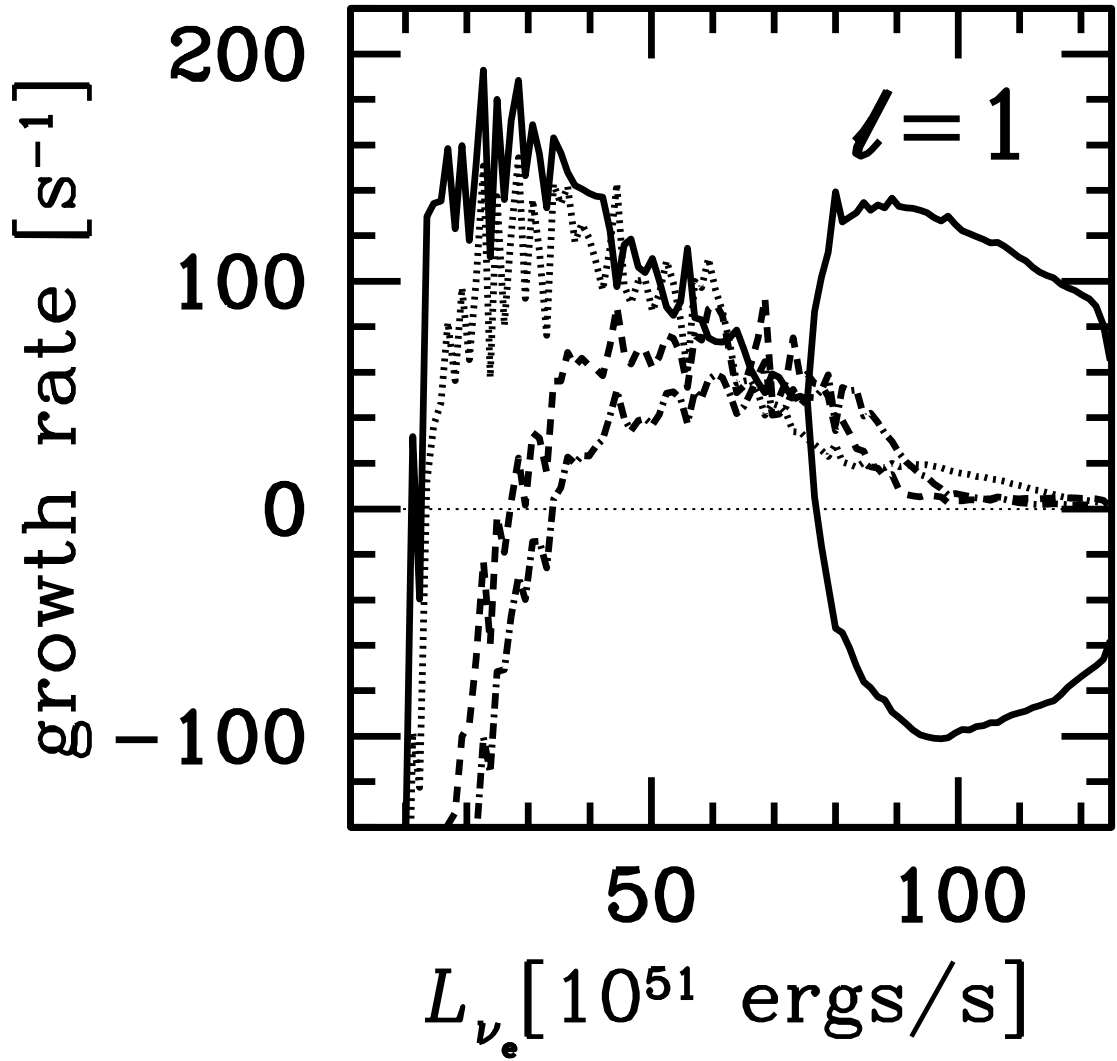}
\caption{The growth rates, ${\rm Re}(\omega)$,  
for $\ell=1$ modes as a function of neutrino luminosity.
The "free" inner boundary condition was imposed for
the radial-velocity perturbation.
The solid curve represents the fundamental mode while the 
dotted, dashed and dash-dotted curves correspond to the first, second
and third overtones, respectively.
\label{fig14}}
\end{figure}

The results are shown in Fig.~\ref{fig13}.
The thermal mode denoted by the long dashed curve is stable
below the crtical luminosity and becomes neutral at the critical point.
The right pannel of the Fig.~\ref{fig13} is a zoom in to the critical point,
and not only the inner solution but also the outer solution is displayed.
They coincide with each other as a neutral mode at the critical luminosity.
An interesting thing is that the outer solutions are also stable
below the critical luminosity,
contrary to the previous results by \citet{yam05}.
It is noted, however, that the spherical accretion flows become
unstable against the non-radial perturbations anyway. 
Another new feature is that the oscillatory radial fundamental mode
(the solid curve in the left panel of Fig.~\ref{fig13})
is not bifurcated up to
the critical luminosity.
The temporal variations of the neutrino sphere somehow delayed
the occurence of the bifurcation.
For the moment, we do not have a good explanation for this phenomenon.
We do not know either whether other non-radial fundamental modes
are also affected or not.

\section{Summary and Discussions}\label{disc}

In this paper, we investigated the stability of the spherically symmetric
accretion flow through the standing shock wave onto the proto neutron star,
which is supposed to approximate the post-bounce situation
in the core-collapse supernova.
We performed systematically the global linear-stability analyses for 
both radial and non-radial perturbations.
We found that the flow is stable for all modes 
if the neutrino luminosity is lower than $\sim 1\times 10^{52}$ ergs/s
in our models with $\dot{M}=1.0M_{\odot}/{\rm s}$.
For larger luminosities, the non-radial instabilities are induced,
probably via the advection-acoustic cycles.
The modes with $\ell=2$ and $3$ become  unstable at first for relatively
low neutrino luminosities, e.g. $\sim 2-3\times 10^{52}$ ergs/s 
for the same accretion rate,
whereas the $\ell=1$ mode is the most unstable for higher luminosities, 
$\sim 3-7\times 10^{52}$ ergs/s.
These are all oscillatory modes. For still higher luminosities, 
$\gtrsim 7\times 10^{52}$ ergs/s, the non-oscillatory modes become unstable,
among which we identified the non-radial ones as convection.
In this way, we discussed the convective instability
in the presence of advection on the same basis.
We found that the growth rates of the convective modes change gradually
with $\ell$ and have a peak at $\ell=5-11$, depending on the luminosity.
The luminosity, at which each convective mode becomes unstable, decreases
with $\ell$ up to $\ell=6$ and then increases for the same accretion rate. 
We confirmed the results obtained by numerical simulations that
the instabilities induced by the advection-acoustic cycles become
more important than convections for lower neutrino luminosities. 
We modified the criterion for the convection
in the presence of advection previously discussed by \citet{fog06a}.
Furthermore, we investigated the sensitivity of the results to
the inner boundary conditions and found that the nature of
the instabilities is not changed qualitatively.

In most of realistic simulations, the neutrino luminosity seems to be
much smaller than the critical value (see e.g. \citep{jan05,sum05}).
Hence, from the results of this paper, the over-stabilizing modes induced
probably by the advection-acoustic cycles are more important than convections 
in causing the global anisotropy as observed in
the ejecta of core-collapse supernovae \citep{leo00,wan02}.
The instability is normally saturated at some amplitudes
owing to the non-linear coupling of various modes~\citep{ohn06a}.
Although the instability of the standing shock wave, whatever the cause, 
is helpful for the shock revival, recent multi-dimensional
numerical simulations \citep{jan05} demonstrated that these instabilities
alone could not induce an explosion. 
Then there must be some other agents to further boost the shock revival.
Very recently \citet{bur06} proposed a new mechanism,
in which the dissipation of out-going acoustic waves that are excited
by g-mode oscillations in the proto neutron star is the agent of the
shock revival.
In this context, the possible coupling between the instabilities of
the accretion flows and the oscillations of the proto neutron star are
interesting~\citep{yos06,ohn06b}.

In this paper, we considered only the spherically symmetric background flows.
Considering the fact that massive stars are rotating in general, we are
naturally concerned with the stability of rotational accretion flows. 
In the previous paper~\citep{yam05}, we showed that the rapid rotation changes 
steady accretion flows in such a way that the critical luminosity
is decreased and, as a result, the shock revival is facilitated.
In so doing, we found that some of the models have
a negative entropy gradient and can be convectively unstable.
>From the results of this paper, we also expect some oscillatory modes will
become unstable and will be more important than convection
for low neutrino luminosities.
Magnetic fields may also play an important role for the stability of
accretion flows.
Then, the so-called magneto-rotational instability (MRI) should be discussed
in the same framework.
In the case of the spherically symmetric background considered in this paper,
each eigen value is degenerate with respect to the index, $m$. 
In the presence of rotation and/or magnetic field, the degeneracy will be
removed at least partially.
Furthermore, how the growth rates themselves are modified by these effects
is an interesting issue.
We are currently undertaking the task~\citep{yam06b} and the results will
be published elsewhere.

\acknowledgments

We are grateful to the refree, T. Foglizzo for his advices including the method
to examine the effects of inner boundary.
This work is partially supported by the Grant-in-Aid for the 21st century
COE program "Holistic Research and Education Center for Physics of
Self-organizing Systems" of Waseda University and for Scientific Research
(14740166, 14079202) of the Ministry of Education, Science, Sports and
Culture of Japan.

\appendix

\section{Heating and Reaction Rates}

As for the neutrino reactions, we take into account only the dominant contributions: the emission and absorption 
by free nucleons.
\begin{equation}
p + e^- \rightleftharpoons \nu_e + n,
\label{eqa1}
\end{equation}
\begin{equation}
n + e^+ \rightleftharpoons \bar{\nu}_e + p.
\label{eqa2}
\end{equation}
Then the heating and reaction rates can be decomposed as
\begin{equation}
\dot{q}=\dot{q}_{\rm ep}-\dot{q}_{\nu{\rm n}}+\dot{q}_{\rm e^+ n}
-\dot{q}_{\bar{\nu}{\rm p}},
\label{eqa3}
\end{equation}
\begin{equation}
\lambda=-\lambda_{\rm ep}+\lambda_{\nu{\rm n}}+\lambda_{\rm e^+ n}
-\lambda_{\bar{\nu}{\rm p}}, 
\label{eqa4}
\end{equation}
and are calculated based on the formulae given by \citet{bru85} as follows.

The absorption and emission kernels for the source terms
in the Boltzmann equation are given as 
\begin{equation}
R^a_1=\frac{1}{\hbar^7 c^7}\frac{G_{\rm F}^2}{\pi}
\eta_{np}(3g_A^2 +g_V^2 )\{1-F_e(E+\Delta)\}(E+\Delta)^2
\sqrt{1-\frac{m_e^2 c^4}{(E+\Delta)^2}},
\label{eqa5}
\end{equation}
\begin{equation}
R^e_1=\exp\{-\beta(E-\mu_{p}-\mu_{e}+\mu_{n})\}\cdot R^a_1,
\label{eqa6}
\end{equation}
where the following notations are used:
\begin{equation}
\eta_{np}=C\int\frac{2d^3p}{(2\pi \hbar c)^3}F_n (E) \cdot (1-F_p (E)),
\label{eqa7}
\end{equation}
\begin{equation}
\Delta=(m_n-m_p)c^2.
\label{eqa8}
\end{equation}
The mass and chemical potential of the particle of type $i$ are
denoted as $m_i$, $\mu_i$, respectively;
$\beta=1/kT$ ($k$ is the Boltzmann constant) is
the inverse temperature and $E=p^2/2m_i$ is the kinetic
energy of each particle; $\hbar$, $c$, $G_{\rm F}$ are
the Planck constant, speed of light and
Fermi coupling constant, respectively; the unit is chosen such a way that 
$G_{\rm F}/(\hbar c)^3 =1.166\cdot 10^{-11} [{\rm MeV}^{-2}]$,
and we adopt the values of form factors as 
$g_V=1$, $g_A=1.23$; $F_i (E)$ is the distribution function
for the particle of type $i$ and is assumed be 
the Fermi-Dirac distribution:
\begin{equation}
F_i (E)=\frac{1}{1+\exp\{\beta(E-\mu_i)\}}.
\label{eqa9}
\end{equation}
In Eq.~(\ref{eqa7}) we have introduced the factor 
\begin{equation}
C=\frac{\displaystyle{n_n}}
{\displaystyle{\frac{2}{(2\pi\hbar)^3}\int F_n(E)d^3p}},
\label{eqa10}
\end{equation}
which takes account of the discrepancy between the realistic EOS for nucleons
we employ in the calculations 
and the ideal-gas EOS assumed in the above formulae.
$n_n$ is the number density of neutron given by the
realistic EOS.

Considering the fact that the region of our interest is optically thin
for neutrinos, that is, outside the neutrino 
sphere, we can write the heating and reaction rates, $\dot{q}_{ep}$, $\dot{q}_{\nu n}$,
$\lambda_{ep}$ and $\lambda_{\nu n}$, as
\begin{equation}
\dot{q}_{ep}=\frac{4\pi c}{(2\pi \hbar c)^3}
\int R^e_1 \{1-f F_{\nu}(E)\}E^3 dE,
\label{eqa11}
\end{equation}
\begin{equation}
\dot{q}_{\nu n}=\frac{4\pi c}{(2\pi \hbar c)^3}
\int R^a_1 f F_{\nu}(E)E^3 dE,
\label{eqa12}
\end{equation}
\begin{equation}
\lambda_{ep}=\frac{4\pi c}{(2\pi \hbar c)^3}
\int R^e_1 \{1-f F_{\nu}(E)\}E^2 dE,
\label{eqa13}
\end{equation}
\begin{equation}
\lambda_{\nu n}=\frac{4\pi c}{(2\pi \hbar c)^3}
\int R^a_1 f F_{\nu}(E) E^2 dE,
\label{eqa14}
\end{equation}
where $fF_{\nu}(E)$ is the local distribution function
of irradiating neutrinos. We take the Fermi-Dirac distribution with a temperature $T_{\nu}$ 
and a vanishing chemical potential for$F_{\nu}(E)$:
\begin{equation}
F_{\nu}(E)=\frac{1}{1+\exp(E/kT_{\nu})},
\label{eqa15}
\end{equation}
whereas $f$ is the so-called geometrical factor,
which takes account of the fraction of solid angle that 
the neutrinos emitted from the neutrino sphere occupy at each point:
\begin{equation}
f=\frac{1-\sqrt{1-(r_{\nu}/r)^2}}{2}.
\label{eqa16}
\end{equation}

In a similar way, the absorption and emission kernels for
the reaction given by Eq.~(\ref{eqa2}) can be written as,
\begin{equation}
R^a_2=\frac{1}{\hbar^7 c^7}\frac{G_{\rm F}^2}{\pi}
\eta_{pn}(3g_A^2 +g_V^2 )\{1-F_{e^+}(E-\Delta)\}(E-\Delta)^2
\sqrt{1-\frac{m_e^2 c^4}{(E-\Delta)^2}}
\Theta(E-\Delta-m_e c^2),
\label{eqa17}
\end{equation}
\begin{equation}
R^e_2=\exp\{-\beta(E-\mu_{n}-\mu_{e^+}+\mu_{p})\}\cdot R^a_2,
\label{eqa18}
\end{equation}
where we use the following notations:
\begin{equation}
\eta_{pn}=C\int\frac{2d^3p}{(2\pi \hbar c)^3}F_p (E) \cdot (1-F_n (E)),
\label{eqa19}
\end{equation}
\begin{equation}
\Theta(x)=\left\{
\begin {array}{rl}
1, &\qquad\mbox{for $x\geq 0$} \\
0, &\qquad\mbox{for $x<0$}.
\end{array} \right.
\label{eqa20}
\end{equation}
Then the heating and reaction rates, $\dot{q}_{e^+ n}$, $\dot{q}_{\bar{\nu}p}$,
$\lambda_{e^+ n}$ and $\lambda_{\bar{\nu}p}$, are given as
\begin{equation}
\dot{q}_{e^+ n}=\frac{4\pi c}{(2\pi \hbar c)^3}
\int R^e_2 \{1-f F_{\bar{\nu}}(E)\}E^3 dE,
\label{eqa21}
\end{equation}
\begin{equation}
\dot{q}_{\bar{\nu} p}=\frac{4\pi c}{(2\pi \hbar c)^3}
\int R^a_2 f F_{\bar{\nu}}(E)E^3 dE,
\label{eqa22}
\end{equation}
\begin{equation}
\lambda_{e^+ n}=\frac{4\pi c}{(2\pi \hbar c)^3}
\int R^e_2 \{1-f F_{\bar{\nu}}(E)\}E^2 dE,
\label{eqa23}
\end{equation}
\begin{equation}
\lambda_{\bar{\nu} p}=\frac{4\pi c}{(2\pi \hbar c)^3}
\int R^a_2 f F_{\bar{\nu}}(E)E^2 dE,
\label{eqa24}
\end{equation}
where $fF_{\bar{\nu}}(E)$ is the local distribution function
of irradiating anti-neutrinos with $F_{\bar{\nu}}(E)$ being 
the Fermi-Dirac distribution function with a temperature $T_{\nu}$ and 
a vanishing chemical potential just like in Eq.~(\ref{eqa16}) and $f$ being 
the geometrical factor.

\clearpage

\clearpage
\clearpage

\end{document}